\documentclass[twoside]{ae100prg}
\usepackage{graphicx}
\usepackage[breaklinks]{hyperref}
\usepackage{booktabs}

\begin{document}
\title{Regular and Chaotic Motion in General Relativity: The Case of a Massive Magnetic Dipole}

\author{Ond\v{r}ej Kop\'a\v{c}ek$^1$, Ji\v{r}\'{i} Kov\'a\v{r}$^2$, Vladim\'{i}r Karas$^1$, Yasufumi Kojima$^3$}
 
\address{$^1$ Astronomical Institute, Academy of Sciences, Bo\v{c}n\'{i}~II~1401/1a, CZ-141\,31~Prague, Czech~Republic}
\address{$^2$ Institute of Physics, Faculty of Philosophy and Science, Silesian University in Opava, Bezru\v{c}ovo n\'{a}m.~13, CZ-746\,01~Opava, Czech~Republic}
\address{$^3$ Department of Physics, Hiroshima University, Higashi-Hiroshima 739-8526, Japan}

\email{kopacek@ig.cas.cz}

\begin{abstract}
Circular motion of particles, dust grains and fluids in the vicinity of compact objects has been investigated as a model for accretion of gaseous and dusty environment. Here we further discuss, within the framework of general relativity, figures of equilibrium of matter under the influence of combined gravitational and large-scale magnetic fields, assuming that the accreted material acquires a small electric charge due to interplay of plasma processes and photoionization. In particular, we employ an exact solution describing the massive magnetic dipole and we identify the regions of stable motion. We also investigate situations when the particle dynamics exhibits the onset of chaos. In order to characterize the measure of chaoticness we employ techniques of Poincar\'{e} surfaces of section and of recurrence plots.
\end{abstract}

\section{Introduction}
This work represents a continuation of our steady effort \cite{kopacek10, kovar10, kovar08} to understand dynamic properties of charged test particles being exposed to the simultaneous action of strong gravitational and electromagnetic fields surrounding compact objects -- neutron stars and black holes. As we bear astrophysical motivation in our mind we choose such fields which could constitute a reasonable model of a real situation occurring in the vicinity of these objects. Survey of the test particle trajectories might be regarded as a single particle approximation to the complex dynamics of the astrophysical plasma which is applicable once the plasma is diluted sufficiently. Regions of diluted plasma are likely to be found above and below the main accretion body of astrophysical systems driven by compact objects.

\begin{figure}[htb]
\centering
\includegraphics[scale=0.28,trim=0mm 0mm 0mm 0mm,clip]{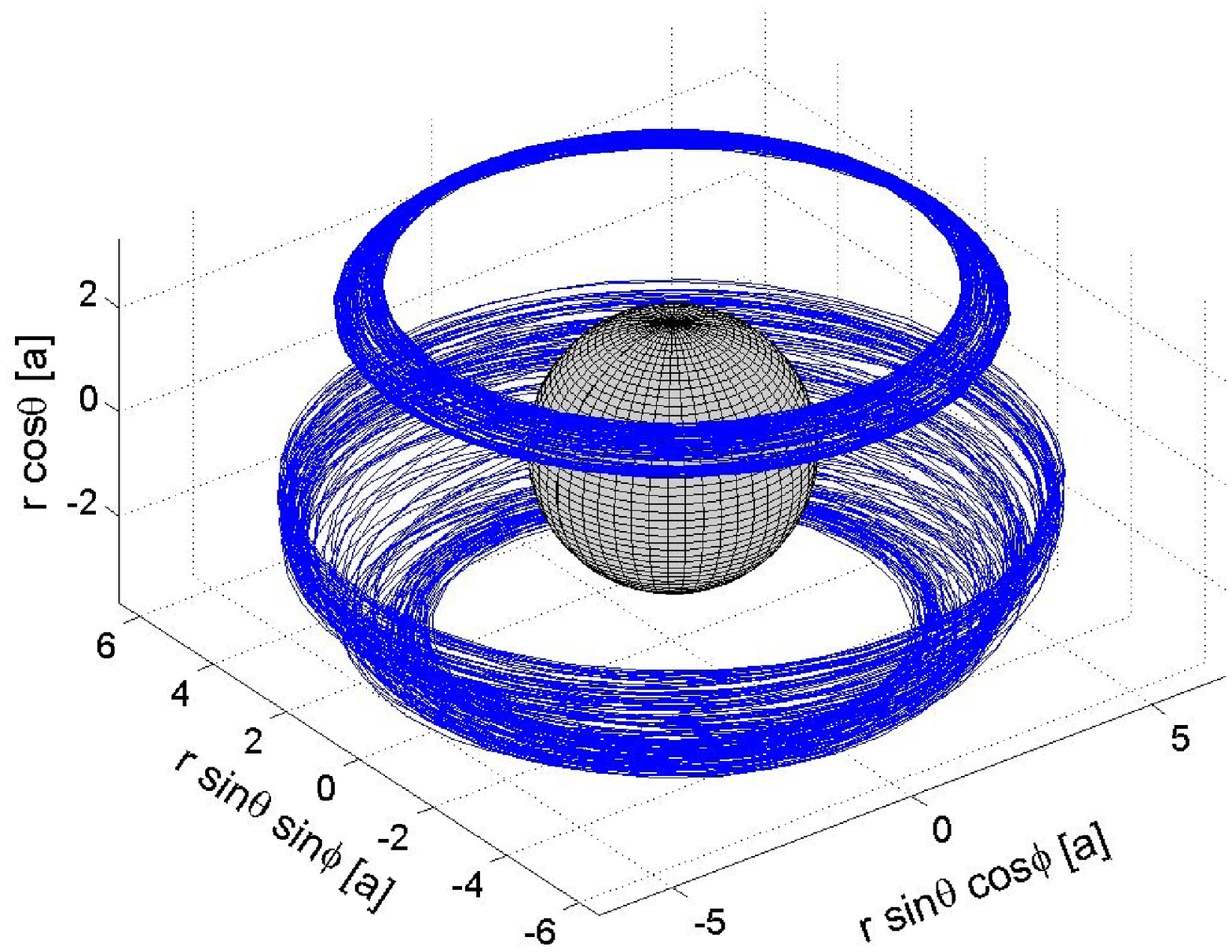}~~
\includegraphics[scale=0.235,trim=0mm 0mm 0mm 0mm,clip]{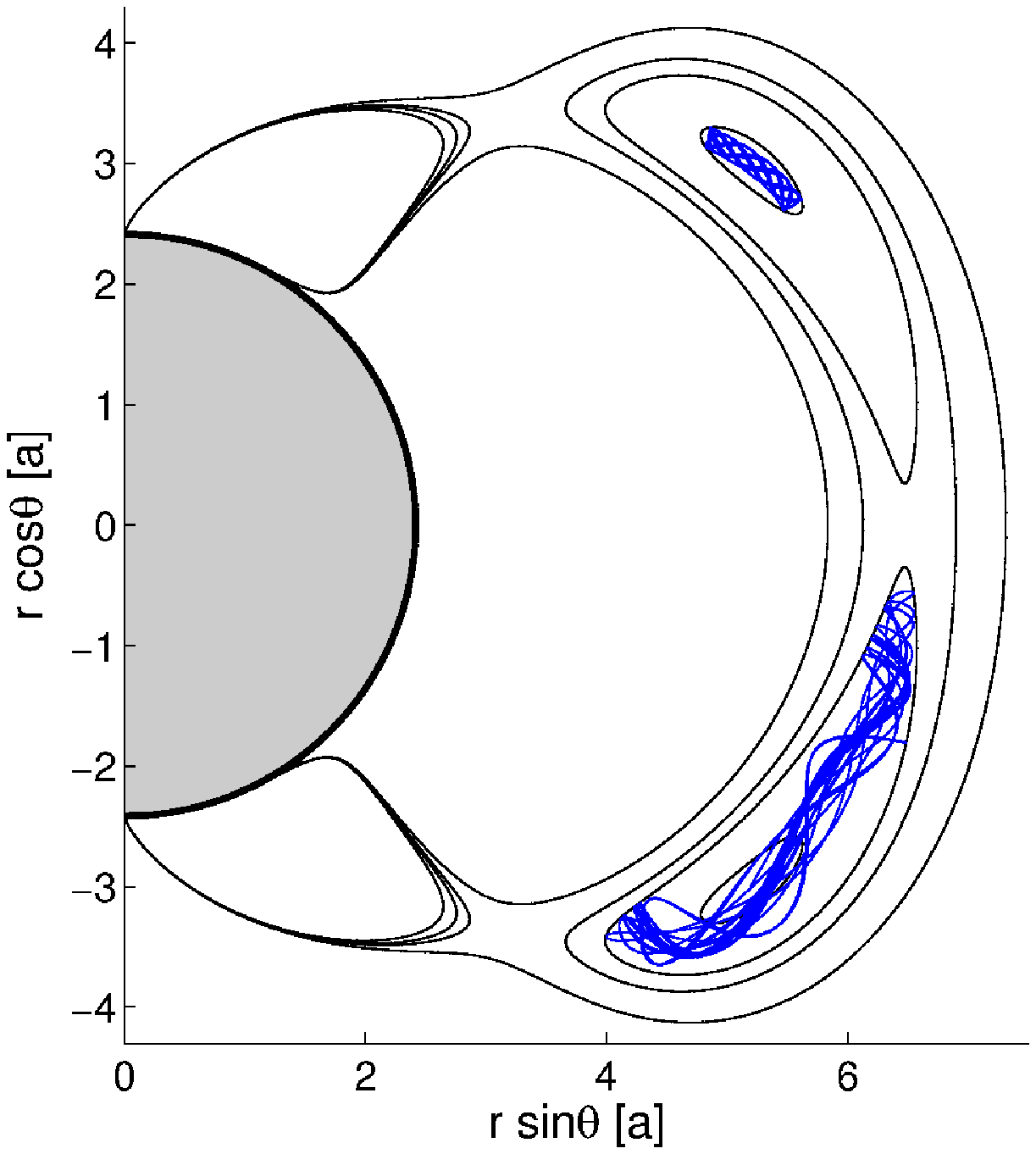}
\caption{Off-equatorial trajectories of charged test particle with $L=-2.356\,a$ and $q=5.581$ in the Bonnor spacetime with $b=1\,a$. In the left panel we present a stereometric projection of two trajectories: the upper one with $E=0.8169$ shows ordered motion while with the higher energy the dynamics acquires properties of deterministic chaos (bottom trajectory with $E=0.8182$). Poloidal projection of these trajectories along with several iso-contours of the effective potential is shown in the right panel. Both particles were launched at $r(0)=6\,a$, $\theta(0)=\pi/3$ with $u^r(0)=0$. Grey color marks $r=r_{\rm h}$ surface in both plots.}
\label{Fig:9}
\end{figure}

In this contribution (which is based mainly on results previously published in \cite{kovar13}) we investigate the motion of the charged test particles around a massive magnetic dipole described by Bonnor's exact solution of coupled Einstein-Maxwell equations \cite{bonnor66}. Such setup allows motion in the off-equatorial lobes if the parameters of the system are chosen carefully. We investigate motion in these lobes. We are particularly curious about the dynamic regime of motion (chaotic versus regular) and how does it change if we alter some of the parameters. Besides the standard technique of Poincar\'{e} surfaces of section we employ the recurrence analysis \cite{marwan07} and show that recurrence plots might serve as an alternative tool to the surfaces of section when analysing individual trajectories.

\section{Massive magnetic dipole}
Using spheroidal coordinates $(t,r,\theta,\phi)$ and geometrized units $c=G=1$ the line element of Bonnor's exact solution \cite{bonnor66} describing the static spacetime around massive magnetic dipole and corresponding vector potential $A_{\alpha}$ are given as follows 
\begin{eqnarray}
{\rm d}s^2&=&-\left(\frac{P}{Y}\right)^2 {\rm d}t^2 +\frac{P^2Y^2}{Q^3Z}({\rm d}r^2+Z{\rm d}\theta^2)+\frac{Y^2Z\sin^2{\theta}}{P^2}{\rm d}\phi^2,\label{metric}\\
A_{\alpha}&=&\left(0,0,0,\frac{2ab r\sin^2{\theta}}{P}\right),
\end{eqnarray}
where $P=r^2-2ar-b^2\cos^2{\theta}$, $Q=(r-a)^2 -(a^2+b^2)\cos^2{\theta}$, $Y=r^2-b^2\cos^2{\theta}$ and $Z=r^2-2ar-b^2$.

The solution is characterized by two independent parameters $a$ and $b$. Inspection of the asymptotic behaviour of the field reveals that these are related to the total mass of the source $M$ as $M=2a$ and to the magnetic dipole moment $\mu$ as $\mu=2ab$. The solution has relatively complicated singular behviour at $P=0$, $Q=0$, $Z=0$ and $Y=0$. However, here we are interested in the regular part of the spacetime only. Therefore we restrict ourselves to $Z>0$ which translates to the condition $r>r_{\rm h} \equiv a+\sqrt{a^2+b^2}$. We investigate the test particle dynamics above the horizon $r_{\rm h}$ only. 

The solution is asymptotically flat (for $a=0$ exactly flat). The metric (\ref{metric}) actually represents a magnetostatic limit of a more general exact solution \cite{pachon06} suggested to describe the exterior field of a rotating neutron star. In the case of Bonnor's solution, the rotation is not considered and the value of a quadrupole mass moment is fixed by values of the parameters $a$ and $b$. Setting $b=0$ reduces the metric to Zipoy-Voorhees metric with $\delta=2$ \cite{zipoy66,voorhees70}.
  
Generalized Hamiltonian (``super Hamiltonian``) describing the motion of the ionised test particle of charge $\tilde{q}$ is given as follows \cite{mtw}:
\begin{equation}
\label{SuperHamiltonian}
\mathcal{H}=\frac{1}{2}g^{\mu\nu}(\pi_{\mu}-\tilde{q}A_{\mu})(\pi_{\nu}-\tilde{q}A_{\nu}),
\end{equation}
where $\pi_{\mu}$ is the generalized (canonical) momentum.

Hamiltonian equations of motion are given in a standard way:
\begin{eqnarray}
\label{HamiltonsEquations}
{\rm d}x^{\mu}/{\rm d}\lambda=\partial \mathcal{H}/\partial \pi_{\mu},\quad d\pi_{\mu}/d\lambda&=&-\partial \mathcal{H}/\partial x^{\mu},
\end{eqnarray}
where $\lambda=\tau/m$ is the affine parameter, $\tau$ the proper time and $m$ represents the rest mass of the particle.

The second Hamilton's equation ensures that the momenta
\begin{eqnarray}
\label{Momenta}
\pi_t&=&p_t+\tilde{q}A_t\equiv-\tilde{E}\\
\label{27}
\pi_{\phi}&=&p_{\phi}+\tilde{q}A_{\phi}\equiv \tilde{L}
\end{eqnarray}
represent constants of motion, reflecting stationarity and axial symmetry of the considered  background.

\begin{figure}[hp!]
\centering
\includegraphics[scale=0.26,trim=0mm 0mm 0mm 0mm,clip]{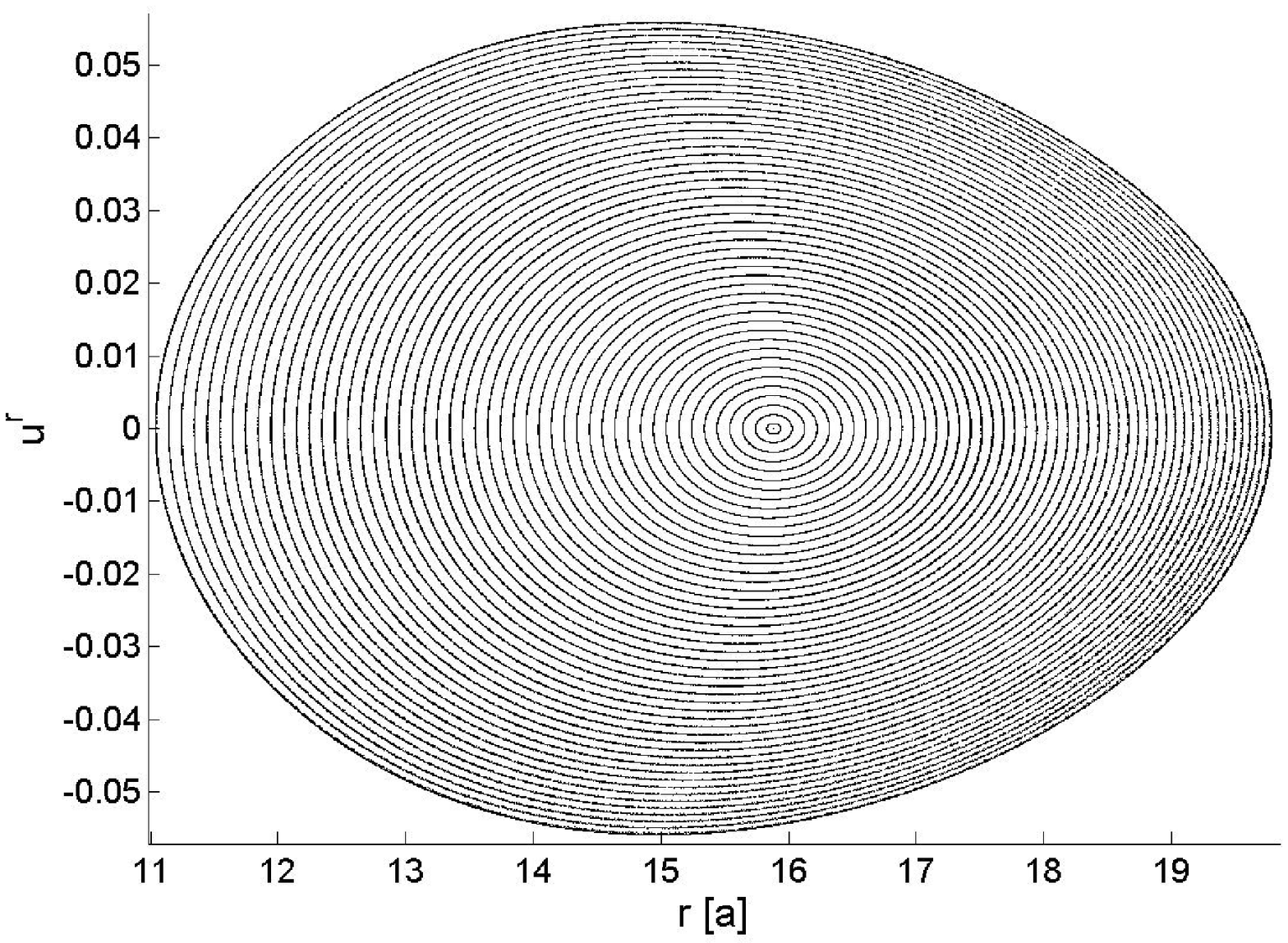}
\includegraphics[scale=0.26,trim=0mm 0mm 0mm 0mm,clip]{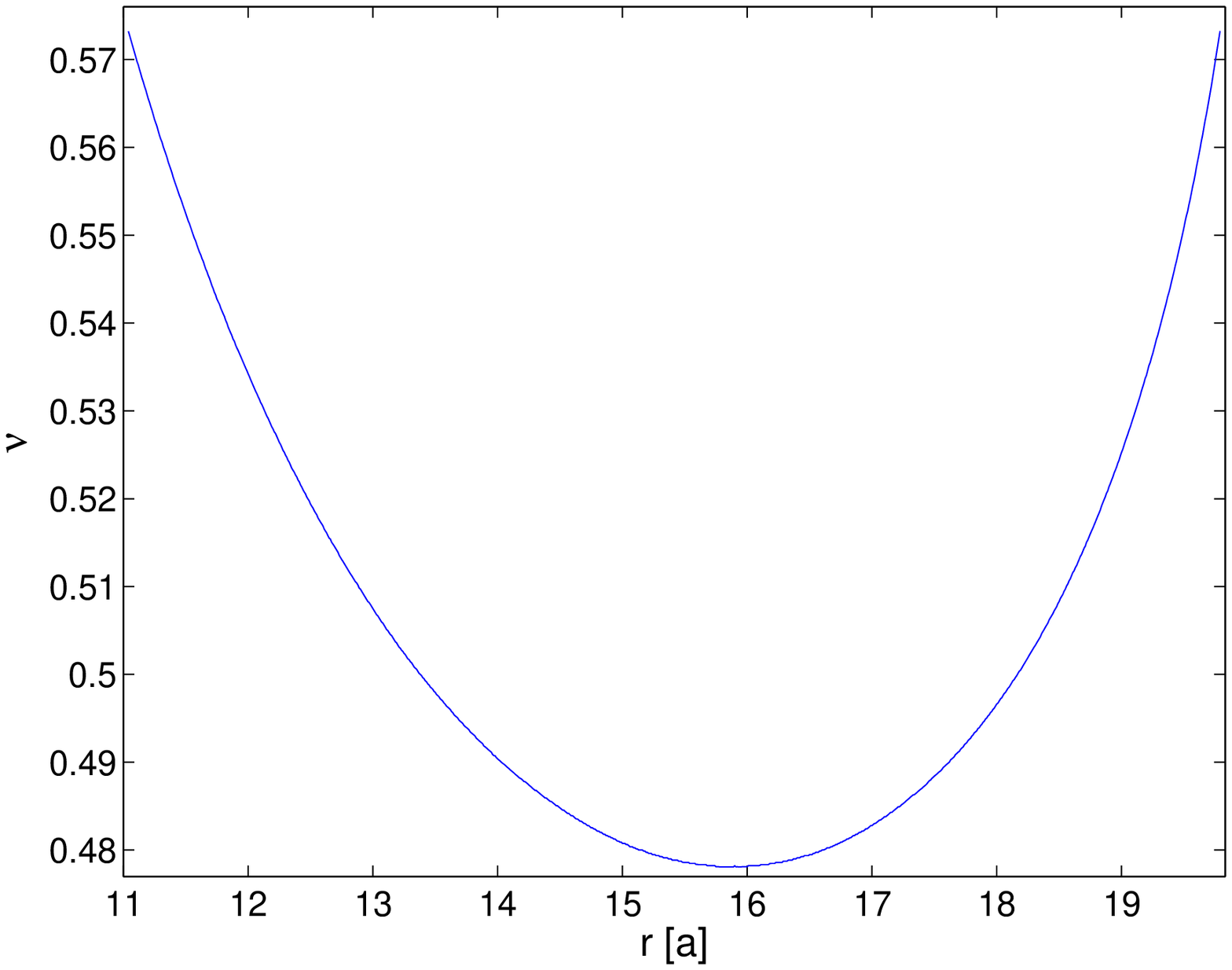}\\
\includegraphics[scale=0.26,trim=0mm 0mm 0mm 0mm,clip]{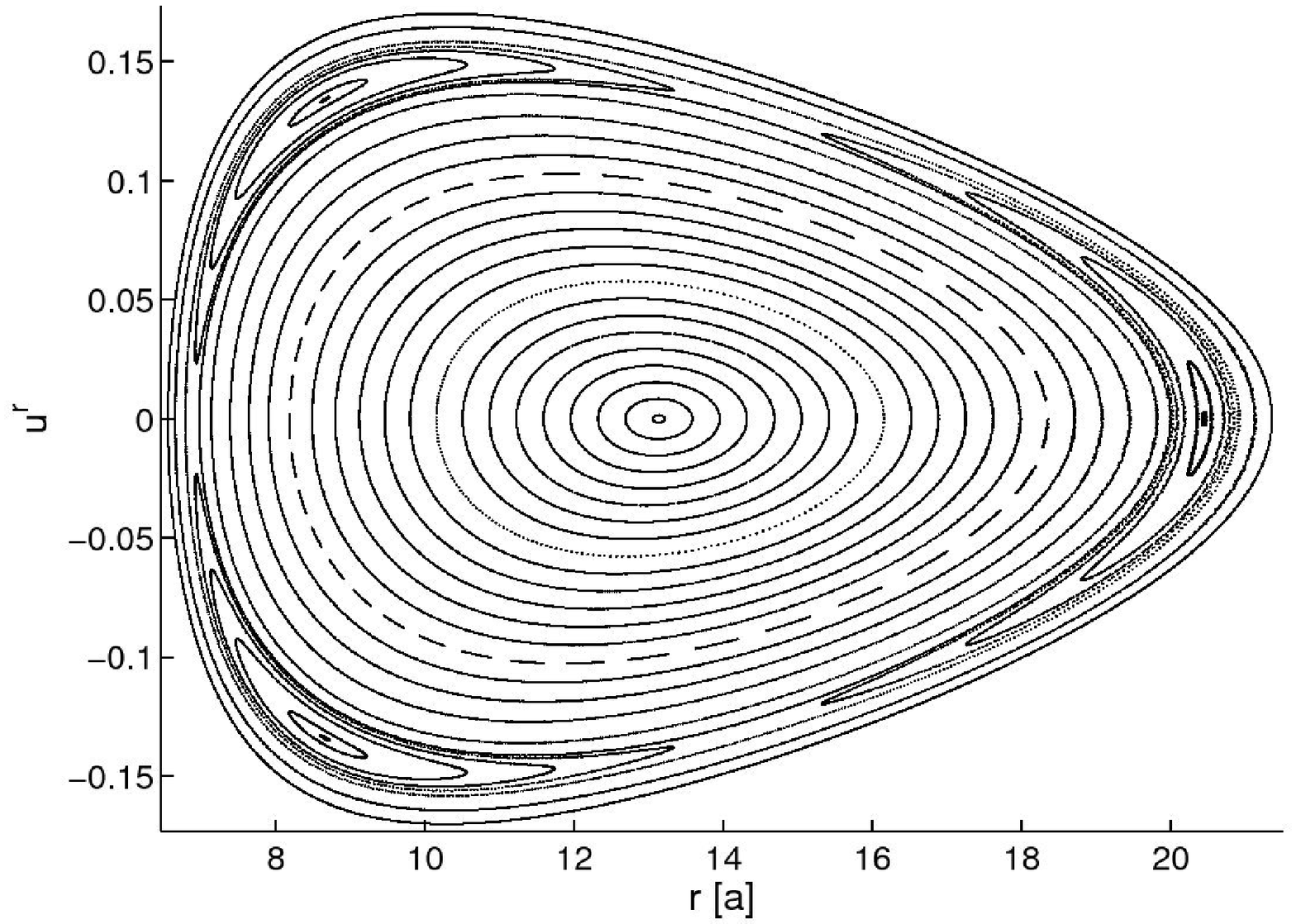}
\includegraphics[scale=0.27,trim=0mm 0mm 0mm 0mm,clip]{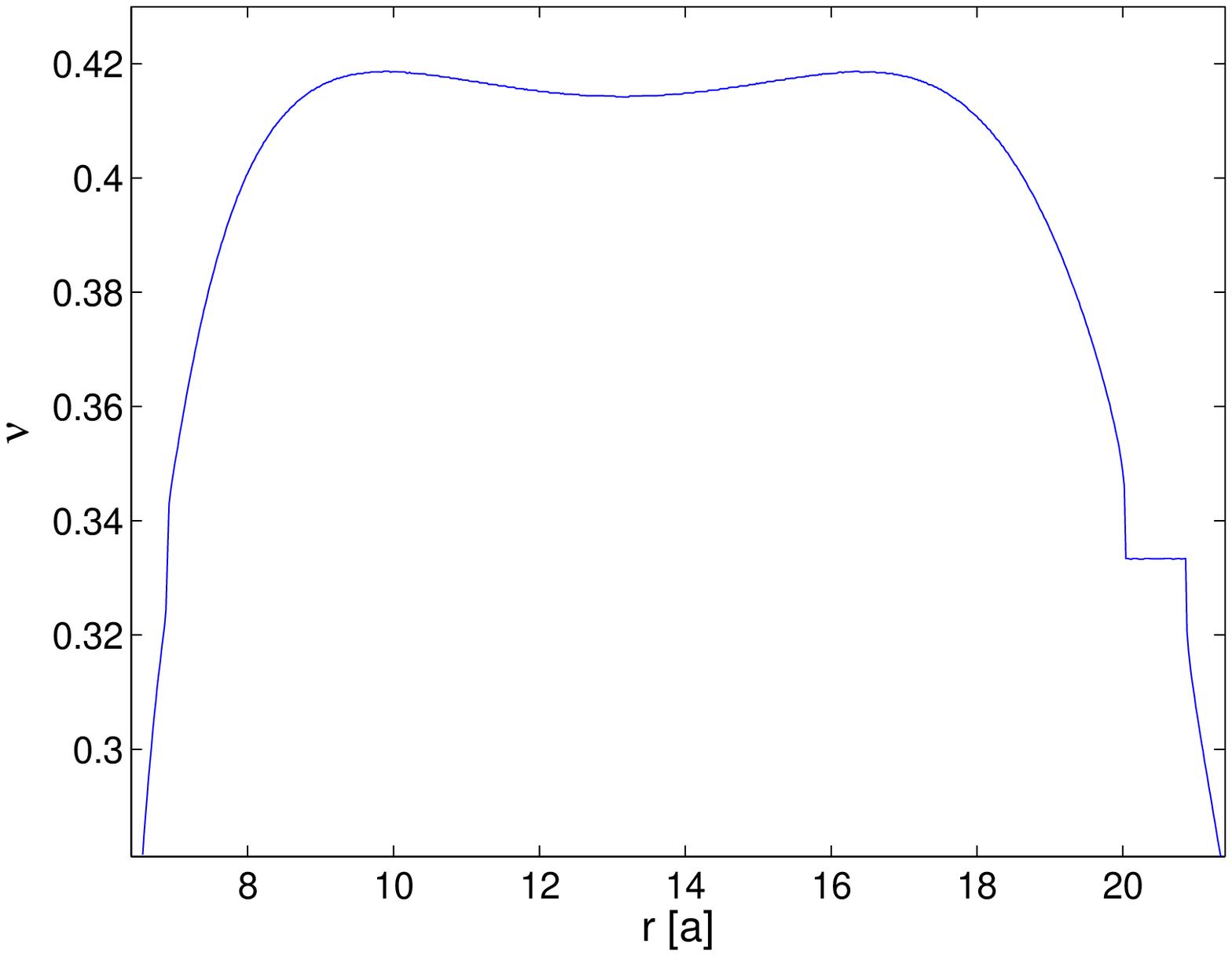}\\
\includegraphics[scale=0.26,trim=0mm 0mm 0mm 0mm,clip]{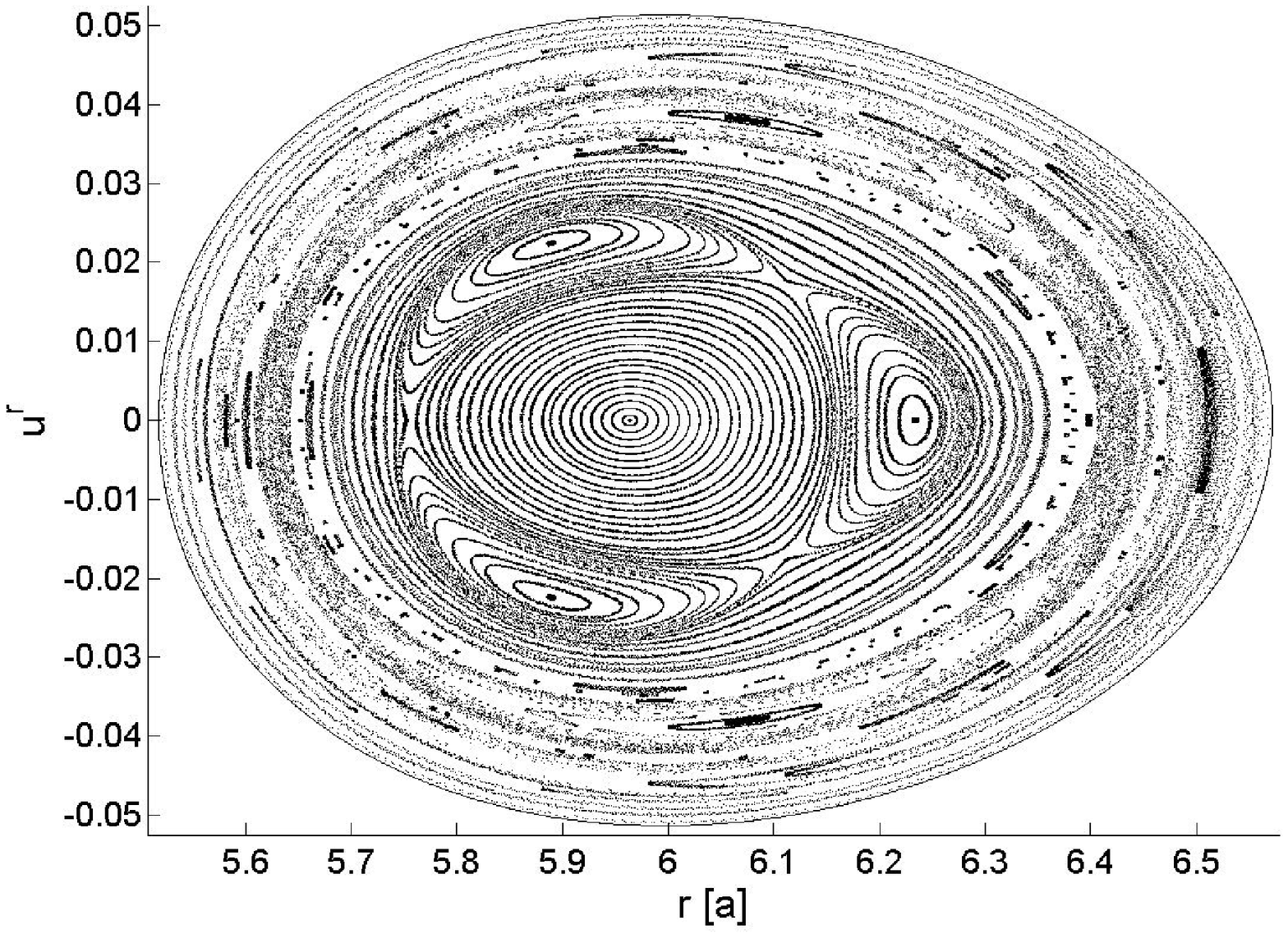}
\includegraphics[scale=0.26,trim=0mm 0mm 0mm 0mm,clip]{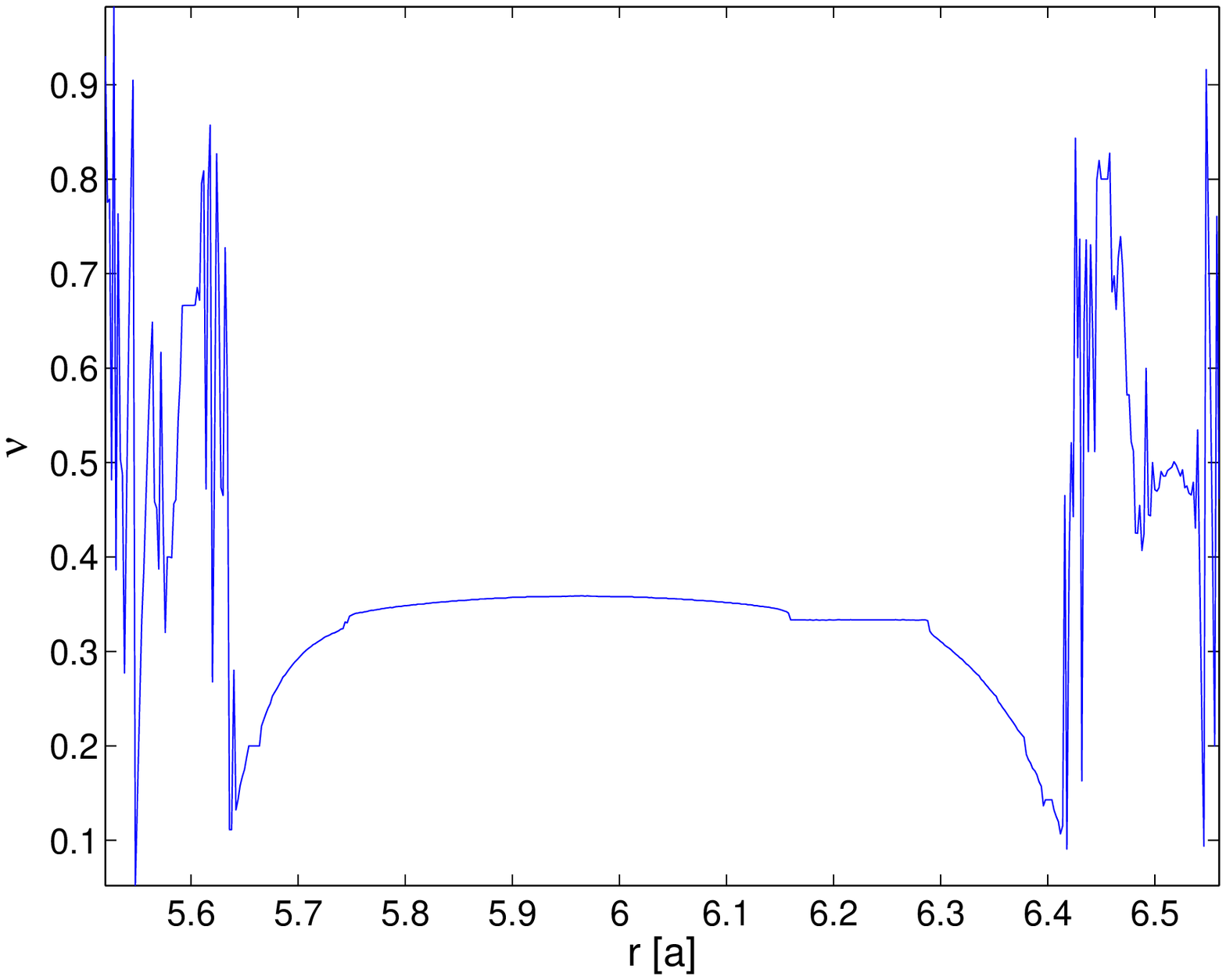}
\caption{Dynamics in equatorial lobes ($\theta_{\rm min}=\theta_{\rm sec}=\pi/2$). Left panels show equatorial surfaces of section while the corresponding rotation curves are shown on the right. Common parameters of the trajectories in the top panels are $E=0.951$, $L=7.2058\,a$, $q=0$ and $b=0$ for which the potential minimum appears at $r_{\rm min}=15\,a$ with $V_{\rm min}=0.9494$. Middle panels show the situation for particles with $E=0.94$, $L=6.1076\,a$, $q=0$ and $b=2.8535\,a$ which brings the stable circular orbit to $r_{\rm min}=10\,a$ with $E_{\rm min}=0.9234$. Bottom panels are plotted for $E=0.818$, $L=-2.7277\,a$, $q=4.7181$ and $b=1\,a$  ($V_{\rm min}=0.8165$ at $r_{\rm min}=6\,a$).}
\label{Fig:10}
\end{figure}

Numerical integration of Hamilton's equations (\ref{HamiltonsEquations}) is carried out using the multistep Adams-Bashforth-Moulton solver of variable order. In several cases when higher accuracy is demanded we employ 7-8th order Dormand-Prince method. Initial values of non-constant components of the canonical momentum $\pi_{r}(0)$ and $\pi_{\theta}(0)$ are obtained from $u^{r}(0)$ (which we set) and $u^{\theta}(0)$ which is calculated from the normalization condition $g^{\mu\nu}u_{\mu} u_{\nu}=-1$ where we always choose the non-negative root as a value of $u^{\theta}(0)$.

\begin{figure}[hp!]
\centering
\includegraphics[scale=0.27,trim=0mm 0mm 0mm 0mm,clip]{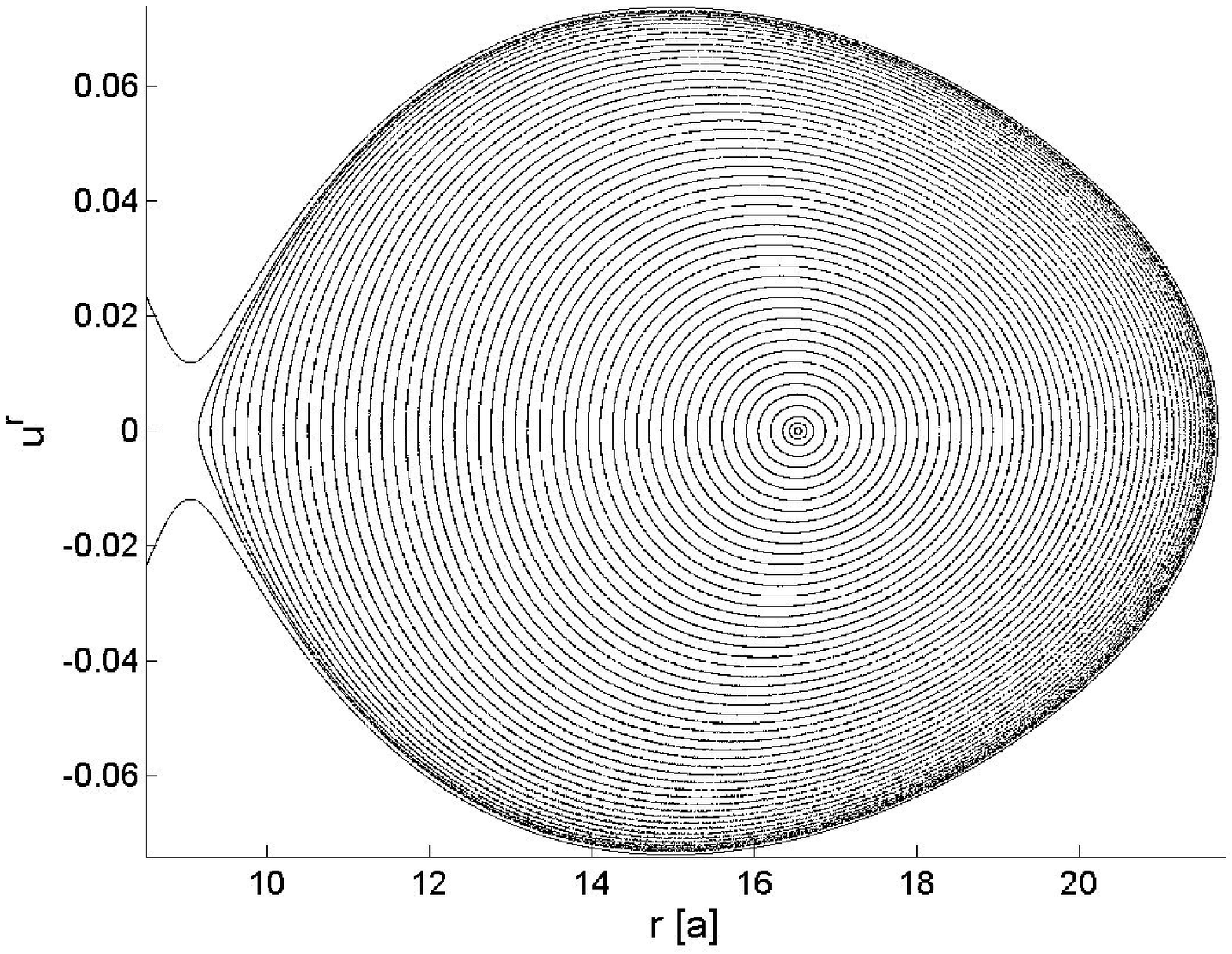}
\includegraphics[scale=0.275,trim=0mm 0mm 0mm 0mm,clip]{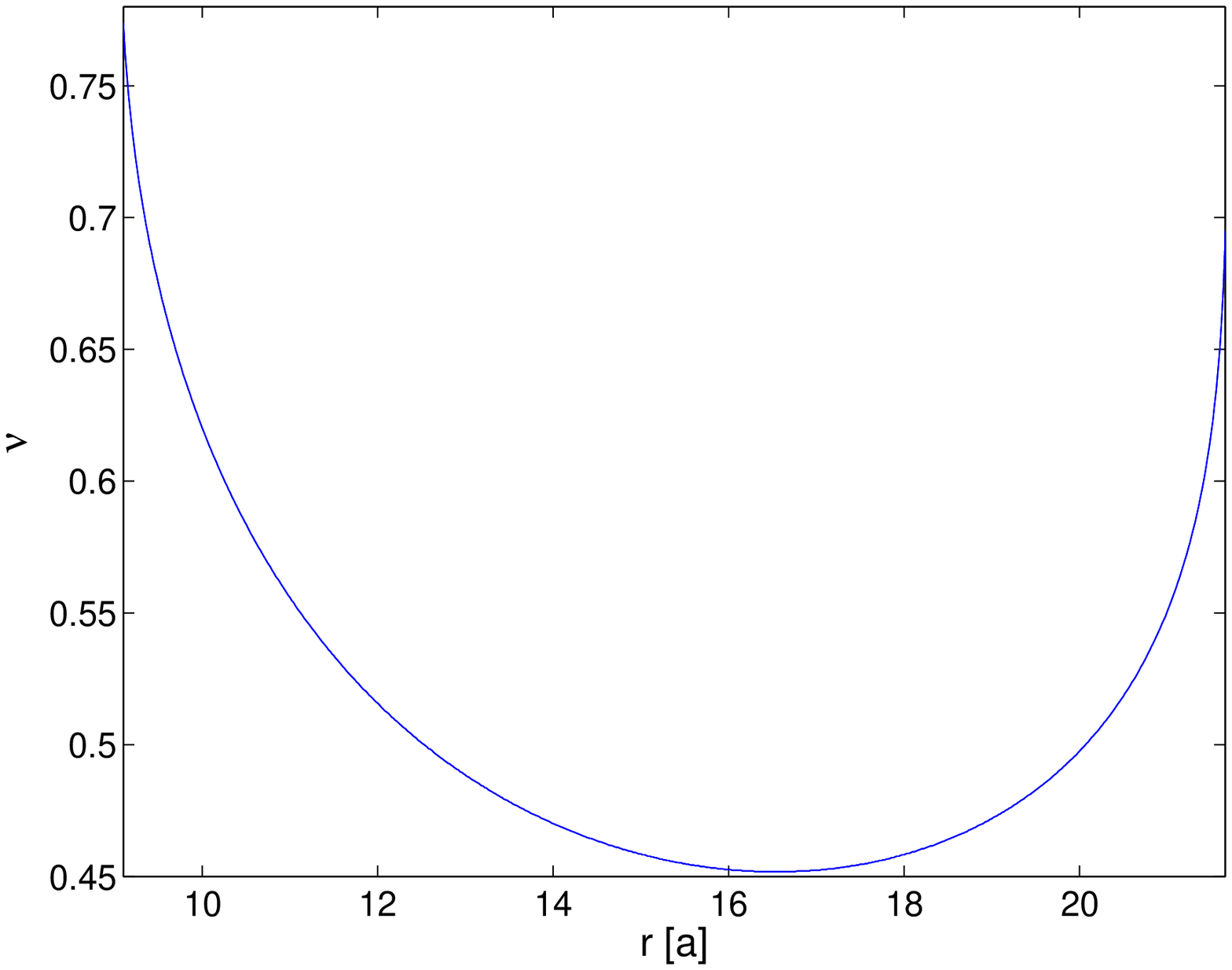}\\
\includegraphics[scale=0.28,trim=0mm 0mm 0mm 0mm,clip]{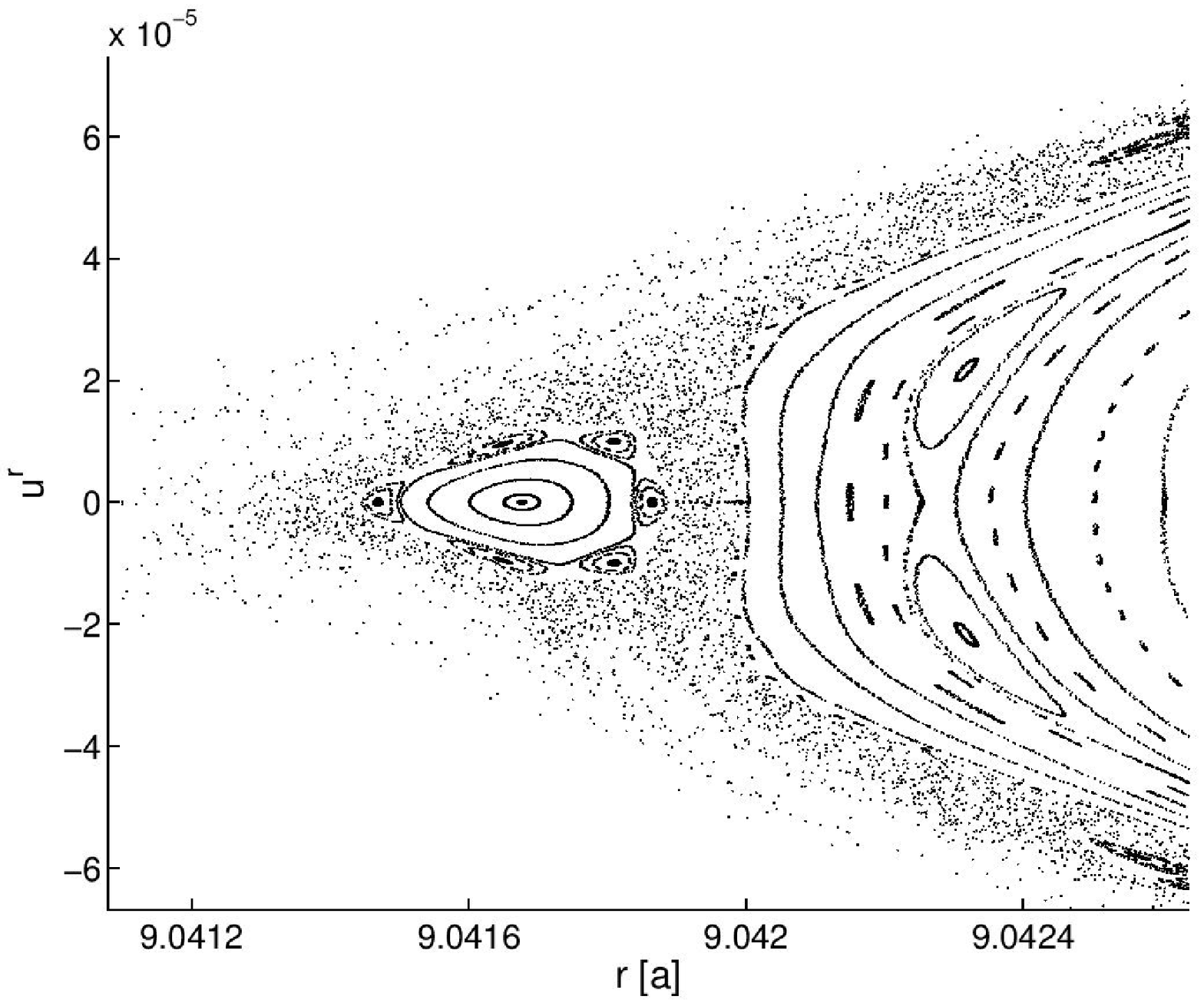}
\includegraphics[scale=0.275,trim=0mm 0mm 0mm 0mm,clip]{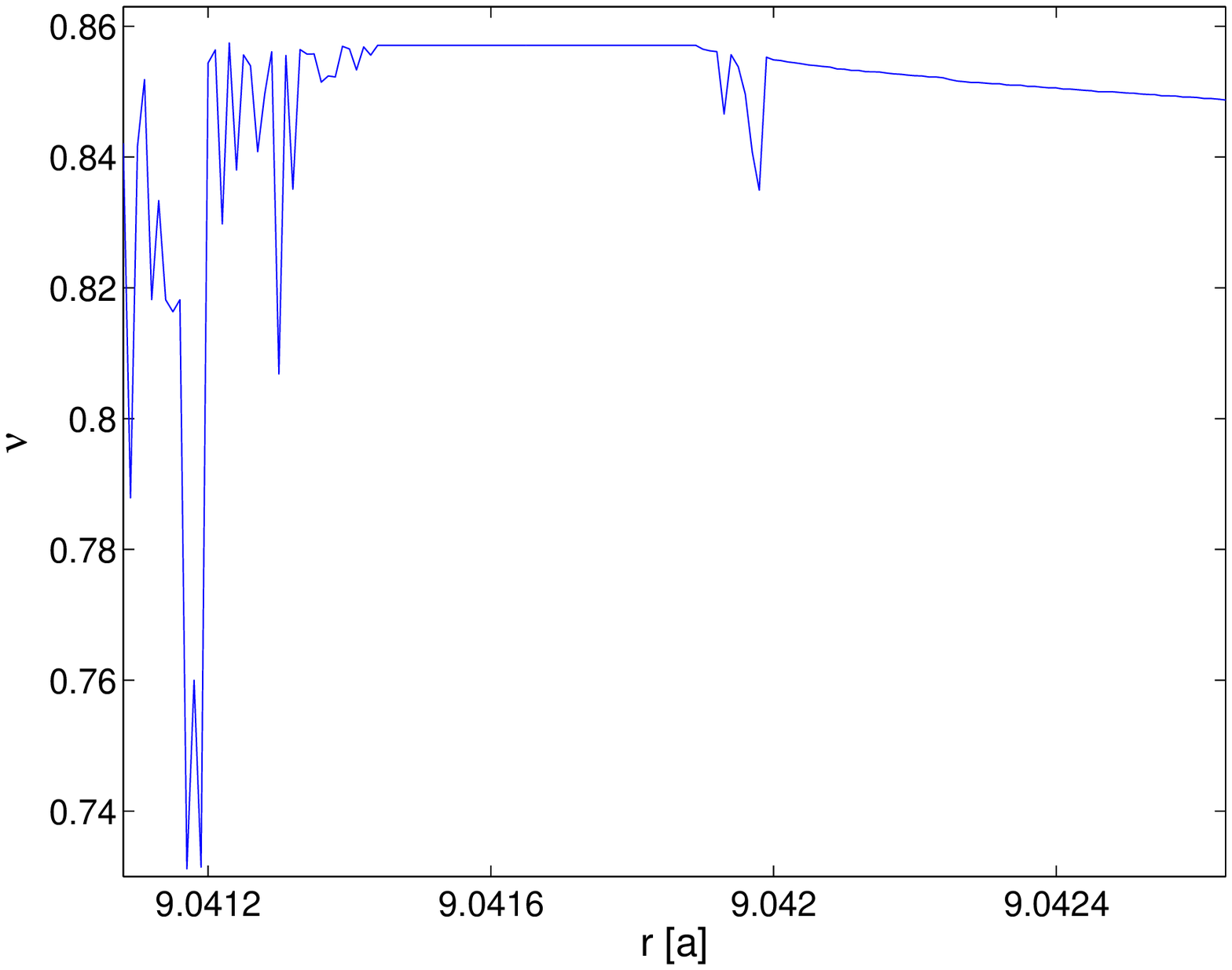}
\caption{Dynamics of test particles in reduced Bonnor spacetime ($b=0$) in the opened equatorial lobe ($\theta_{\rm min}=\theta_{\rm sec}=\pi/2$). Upper panels show regular dynamics, however, zooming the region near the throat reveals the presence of chaotic orbits (bottom panels). Common parameters of the trajectories are $E=0.9522$, $L=7.2058\,a$, $q=0$ and $b=0$. Unlike the case shown in the upper panels of the figure \ref{Fig:10} here the potential lobe is opened allowing the particles to fall onto the horizon. Stability island observed in the bottom surface of section corresponds to $\nu=6/7$ resonance.}
\label{Fig:10b}
\end{figure}

Two-dimensional (i.e. related to the motion in two coordinates, $r$ and $\theta$) effective potential may be expressed as follows:
\begin{equation}
\label{EffectiveNS}
V_{\mathrm{eff}}^2=\frac{P^2}{Y^2} \left[1 + \frac{P^2}{Y^2Z\sin^2{\theta}}
\left( L-q\mu\frac{r\sin^2{\theta}}{P} \right)^2 \right],
\end{equation}
where we introduce specific angular momentum $L\equiv \tilde{L}/m$ and specific charge $q\equiv \tilde{q}/m$.


\section{Dynamics of test particles}

We perform a numerical survey of dynamics of test particles moving within potential wells formed along both the equatorial and off-equatorial circular orbits (so called halo orbits, illustrated by figure \ref{Fig:9}). In order to do so, we apply several complementary methods of investigation of nonlinear dynamic systems. First of all, we construct Poincar\'{e} surfaces of section which give an overall perspective of the phase space dynamics on a given energy hypersurface (for a given values of system parameters). For the inspection of individual trajectories, however, we prefer to analyse their recurrence plots \cite{marwan07}, which proved to be very useful method in our previous work \cite{kopacek10}. Besides other properties of recurrence plots (RPs), we highlight their ability to clearly distinguish between chaotic and regular dynamics on a short time scale, thus reducing the integration time needed for the analysis. We also investigate intrinsic frequencies of the orbits employing the rotation number $\nu$ \cite{contopoulos02} that allows us to detect and locate resonances of the system.

\begin{figure}[ht]
\centering
\includegraphics[scale=0.28,trim=0mm 0mm 0mm 0mm,clip]{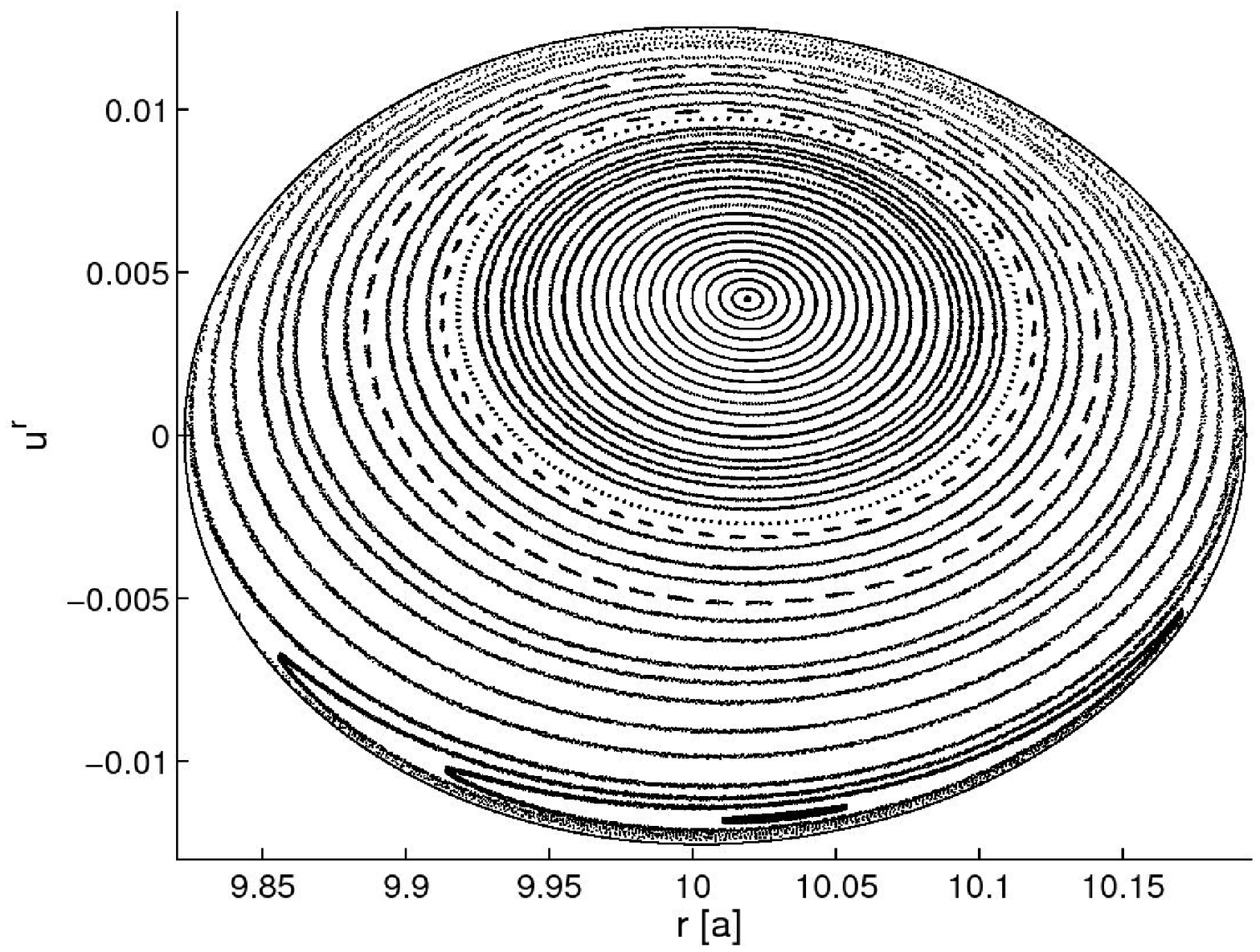}
\includegraphics[scale=0.28,trim=0mm 0mm 0mm 0mm,clip]{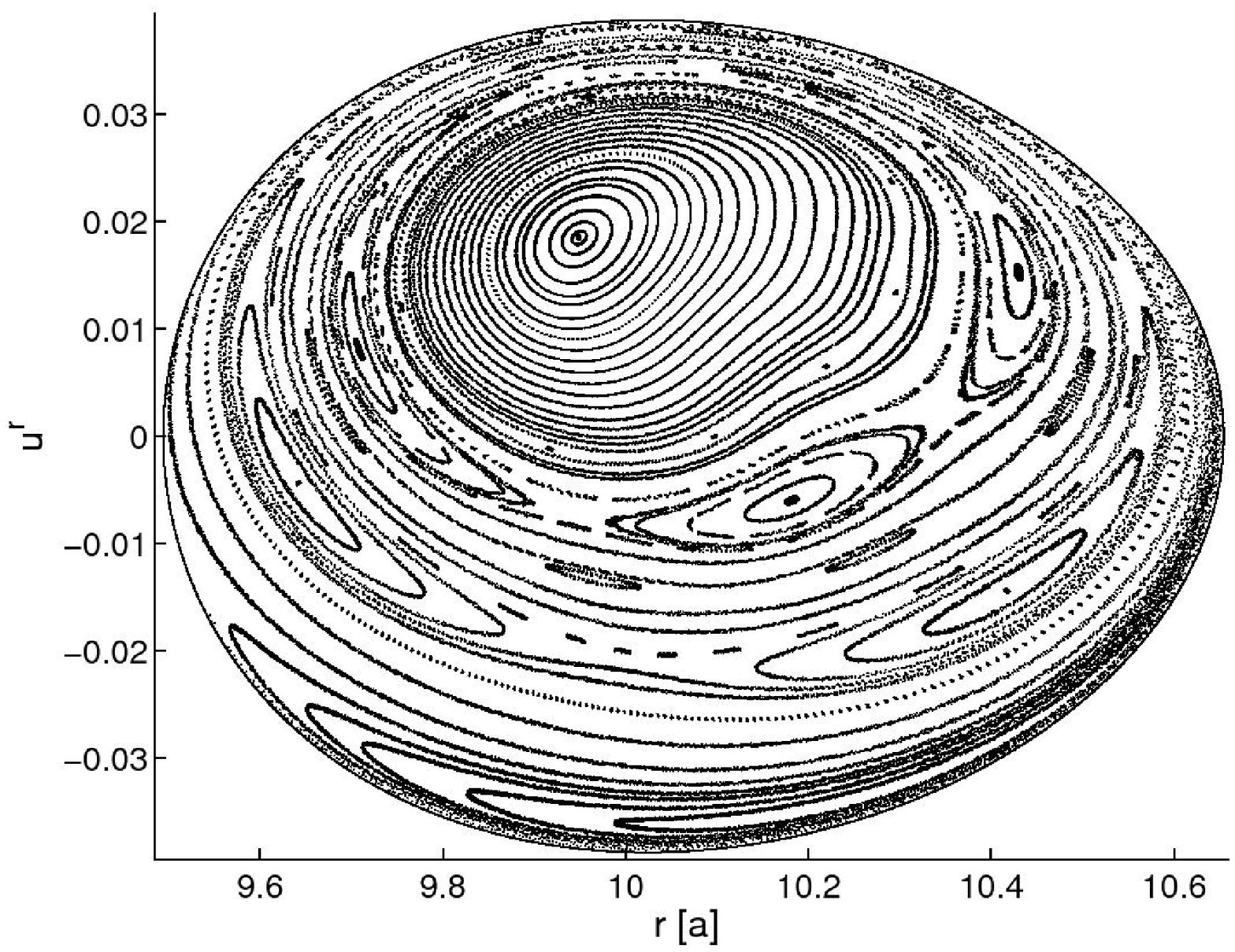}
\includegraphics[scale=0.28,trim=0mm 0mm 0mm 0mm,clip]{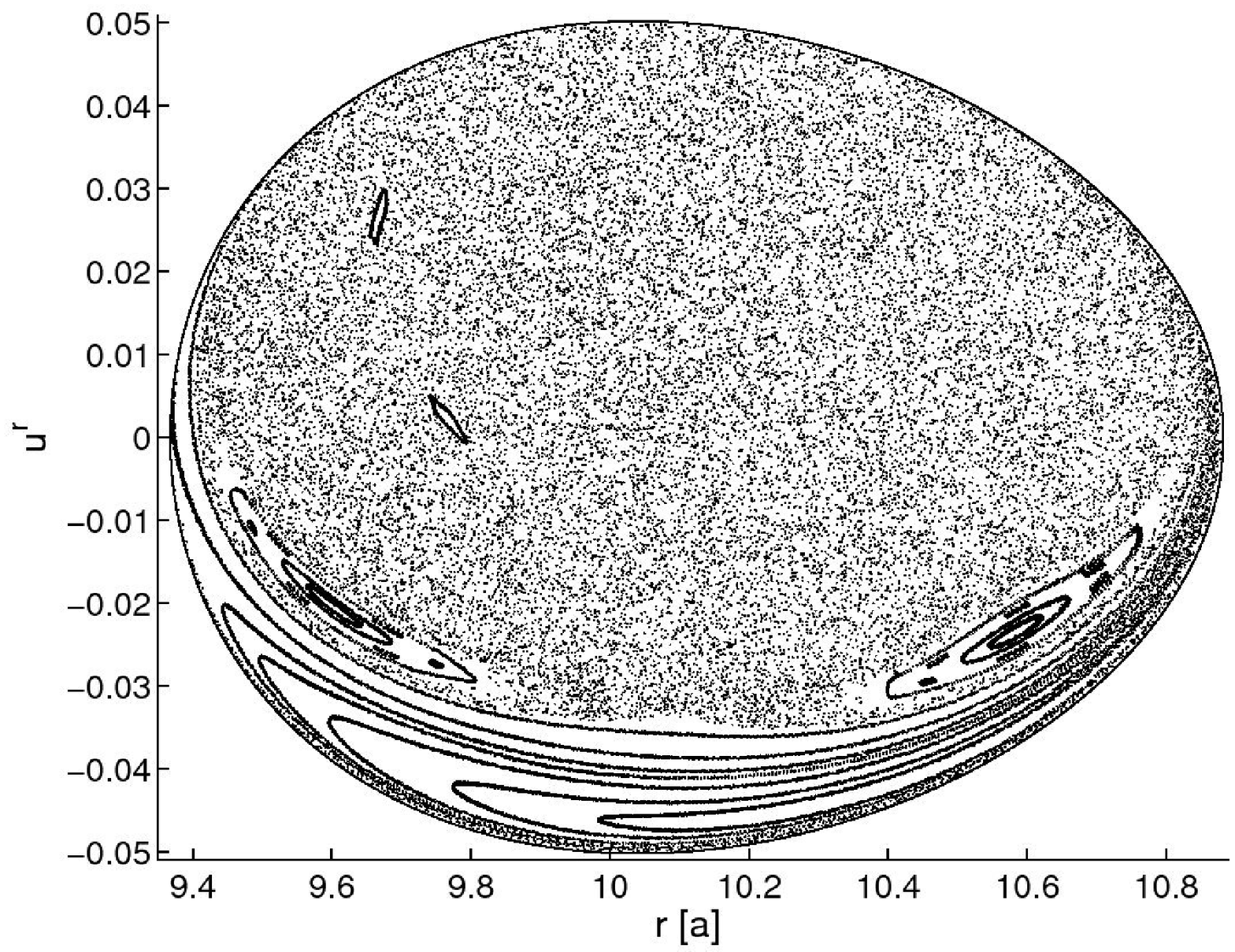}
\includegraphics[scale=0.28,trim=0mm 0mm 0mm 0mm,clip]{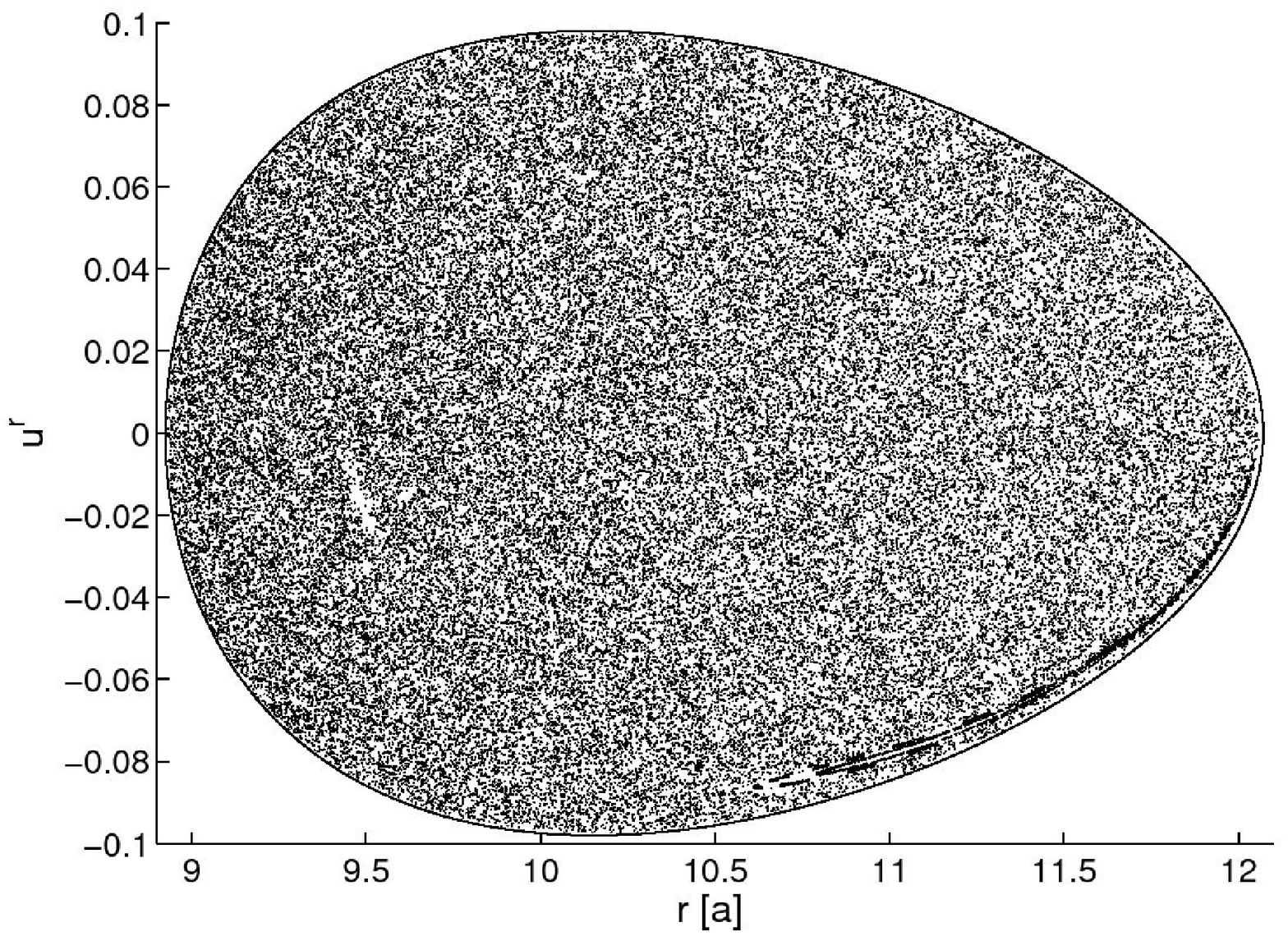}
\caption{Poincar\'{e} surfaces ($\theta_{\rm sec}=\pi/3$) for uncharged particles moving within halo lobes in the Bonnor spacetime with $b=5.9771\,a$. Particles with $L=3.6743\,a$ are launched from the vicinity of the potential minimum ($r_{\rm min}=10\,a$, $\theta_{\rm min}=\theta_{\rm sec}=\pi/3$ and $V_{\rm min}=0.8717$) with various values of energy. The upper left panel shows the section for the level $E=0.8718$ (small halo lobe), in the upper right we set $E=0.873$ (large halo lobe), $E=0.8739$ produces cross-equatorial lobe which just emerged from symmetric halo lobes (bottom left panel), while with $E=0.88$ we obtain the large cross-equatorial lobe which almost opens.}
\label{Fig:11}
\end{figure}


\begin{figure}[ht]
\centering
\includegraphics[scale=0.28,trim=0mm 0mm 0mm 0mm,clip]{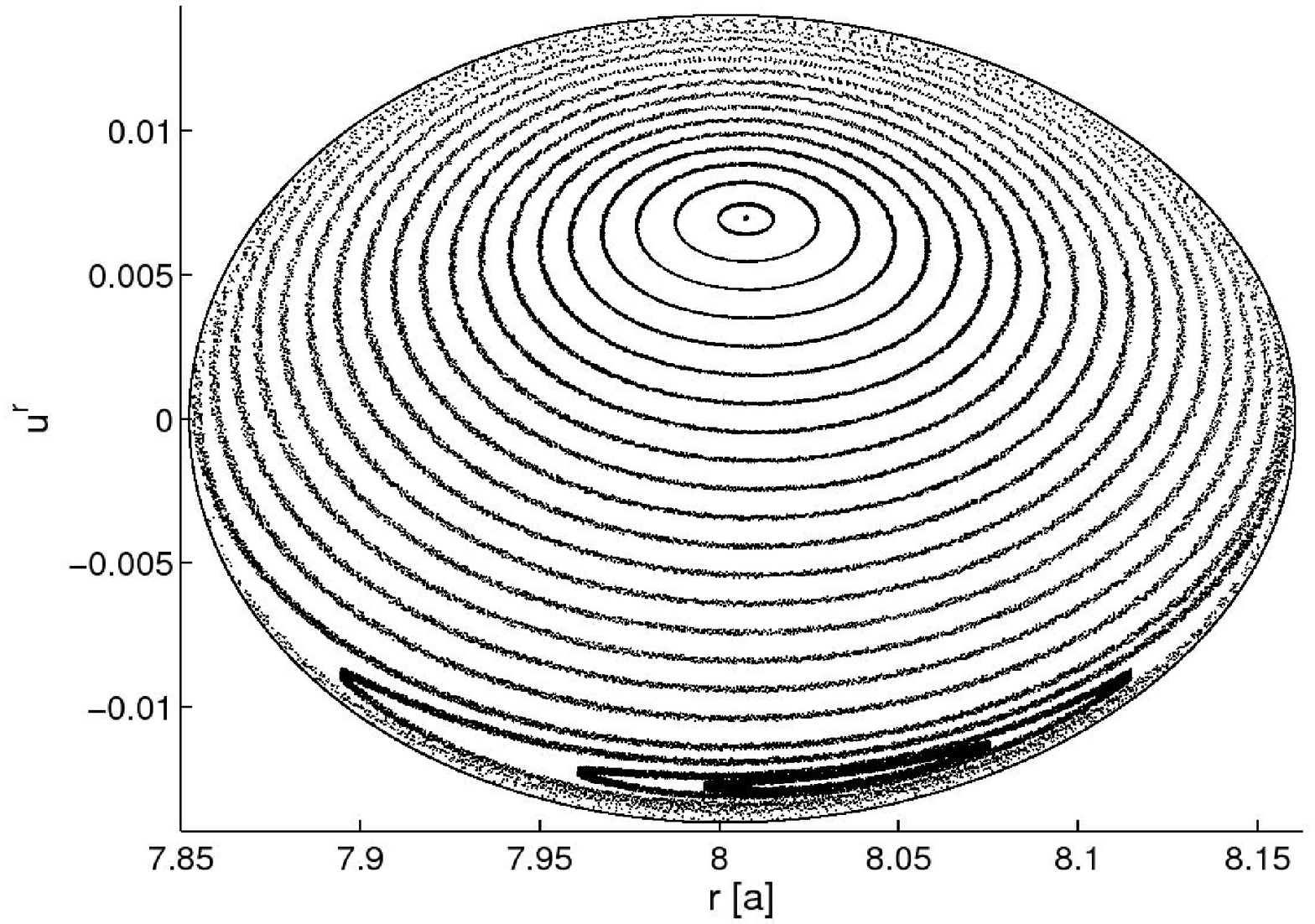}
\includegraphics[scale=0.28,trim=0mm 0mm 0mm 0mm,clip]{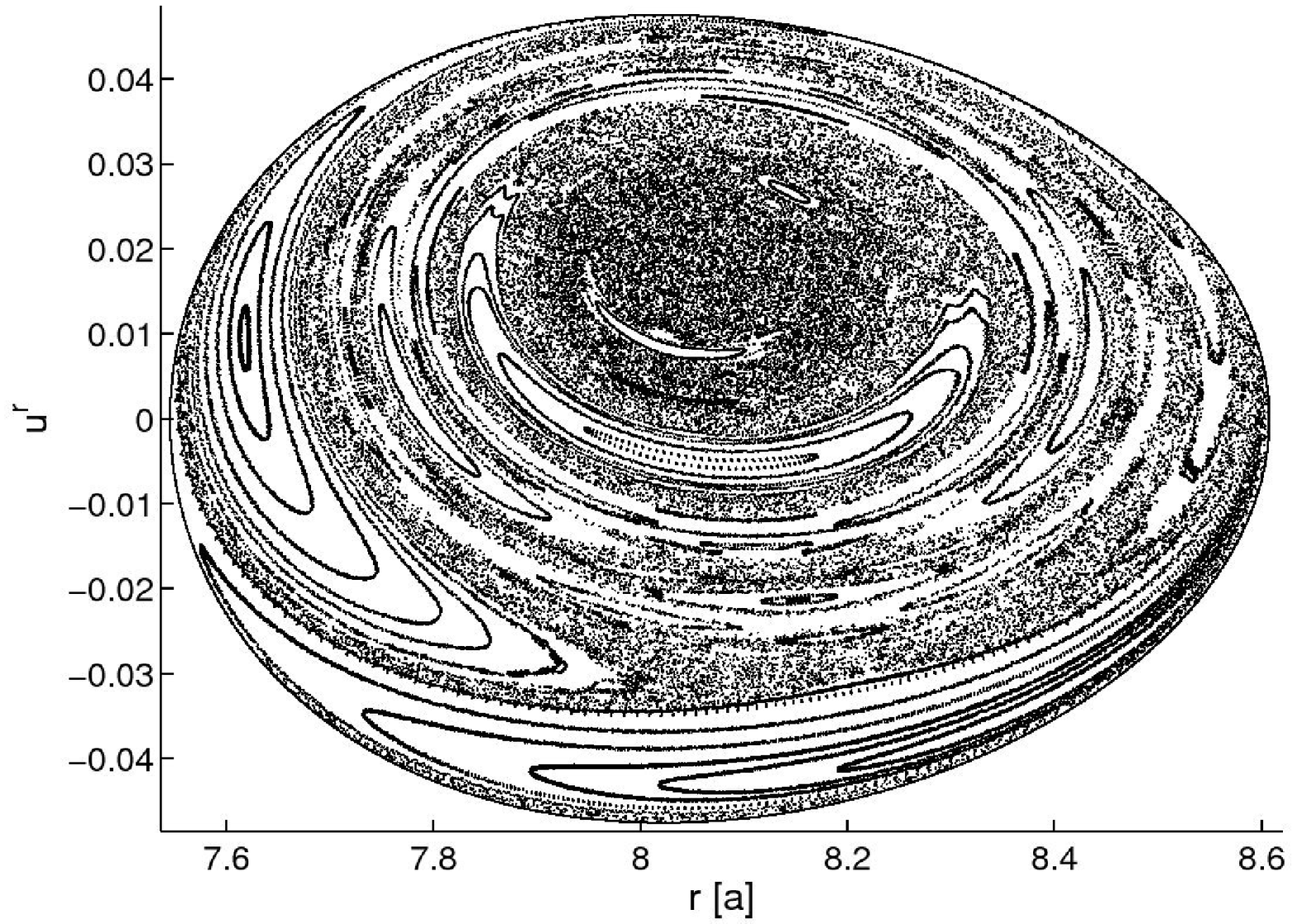}
\includegraphics[scale=0.28,trim=0mm 0mm 0mm 0mm,clip]{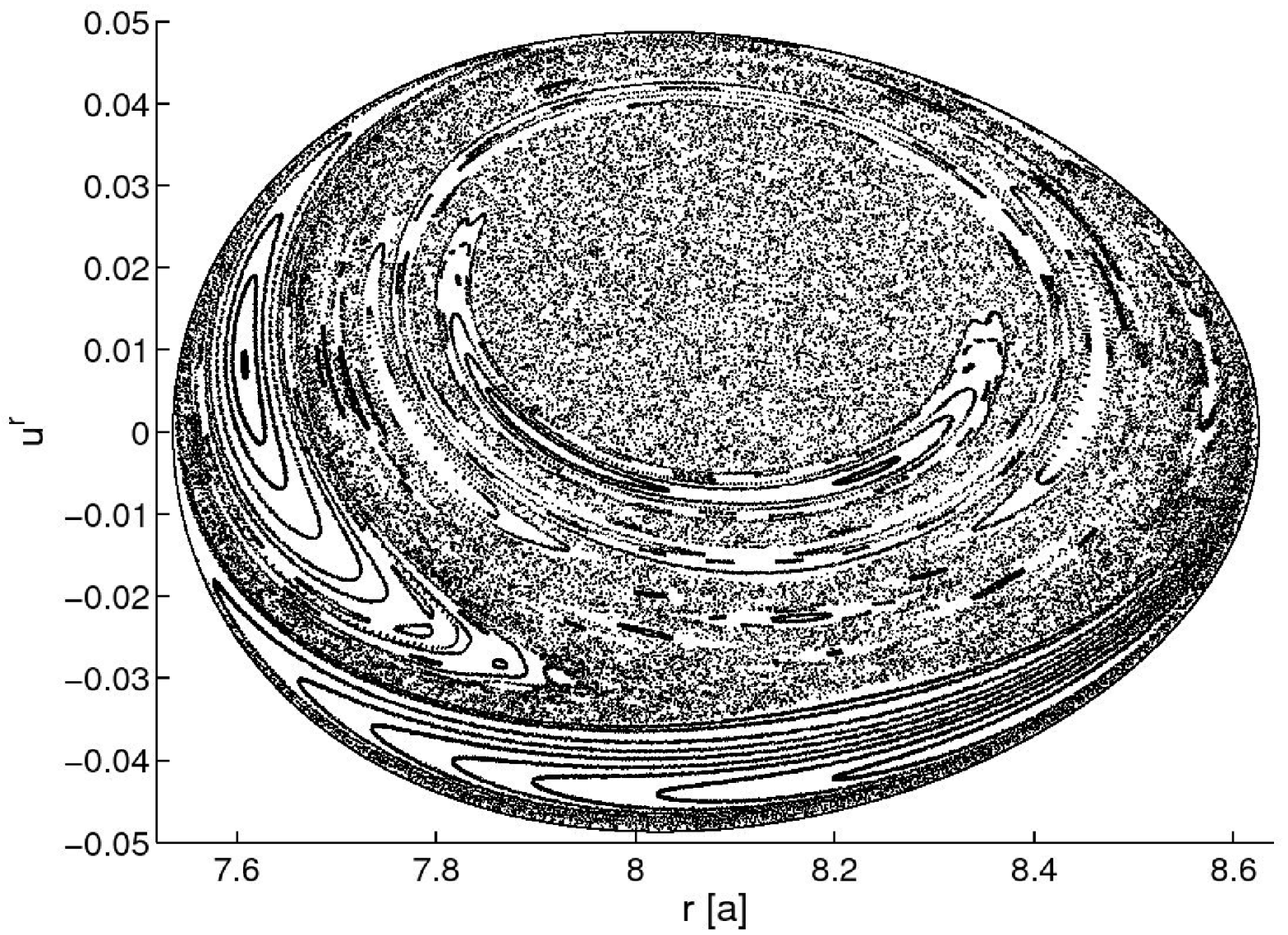}
\includegraphics[scale=0.28,trim=0mm 0mm 0mm 0mm,clip]{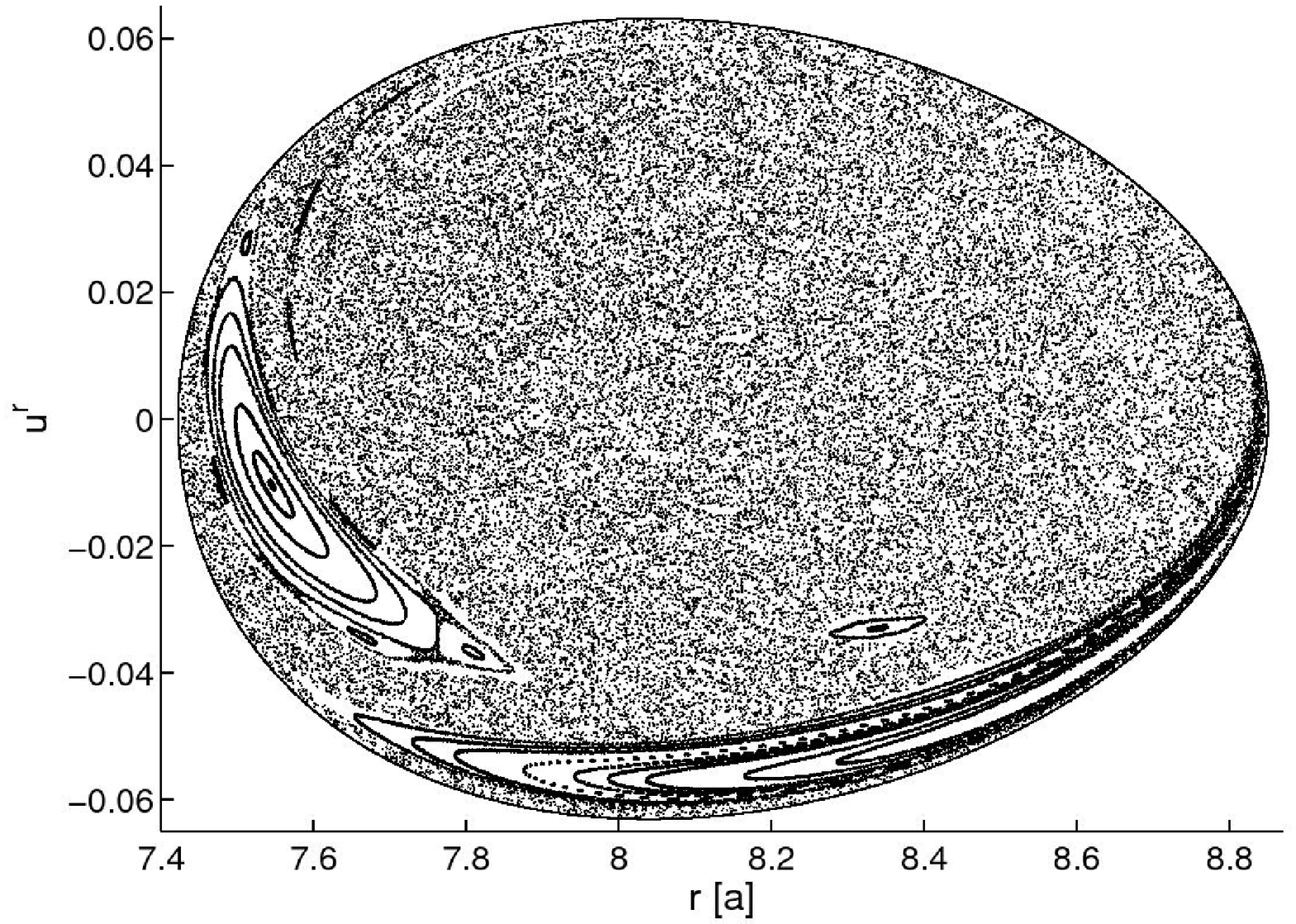}
\caption{Poincar\'{e} surfaces of section ($\theta_{\rm sec}=\pi/3$) for charged particles with $q=0.1259$ moving within halo lobes in the Bonnor spacetime with $b=4.5393\,a$. Particles with $L=-3.5486\,a$ are launched from a vicinity of the potential minimum ($r_{\rm min}=8\,a$, $\theta_{\rm min}=\theta_{\rm sec}=\pi/3$ and $V_{\rm min}=0.8475$) with various values of energy. The upper left panel shows the section for the level $E=0.8477$ (small halo lobe), in the upper right we set $E=0.8495$ (large halo lobe), $E=0.8496$ produces cross-equatorial lobe which just emerged from symmetric halo lobes (bottom left panel) while with $E=0.851$ we obtain the large cross-equatorial lobe.}
\label{Fig:12}
\end{figure}


In order to compare dynamic properties of the system in all three cases (particle in a non-magnetized $b=0$ spacetime, uncharged particle in a magnetized spacetime, and a general case $q\neq0$, $b\neq0$), we first investigate the motion in equatorial potential wells, since there are no circular halo orbits for $b=0$. In figure \ref{Fig:10}, we present series of Poincar\'e surfaces of section with $\theta_{\rm sec}=\pi/2$ along with the corresponding plots of a rotation number as a function of initial value of radial coordinate. 

The upper left panel of figure \ref{Fig:10} shows the Poincar\'{e} surface of the test particles trajectories when the magnetic dipole is switched off by setting $b=0$. We observe perfectly ordered motion with no traces of secondary fixed points nested in Birkhoff islands nor the chaotic orbits. Such a simple pattern on the section is characteristic for integrable systems. The integrability conjecture is also supported by the behaviour of rotation number, which is smooth and non-constant throughout the lobe. While we show roughly $50$ trajectories on the section, the $\nu$-plot is constructed from around $1000$ of them, which allows much more detailed inspection for the possible presence of tiny chaotic domains or faint resonances. 

\begin{figure}[hp!]
\begin{center}
\includegraphics[scale=0.18,trim=0mm 0mm 0mm 0mm,clip]{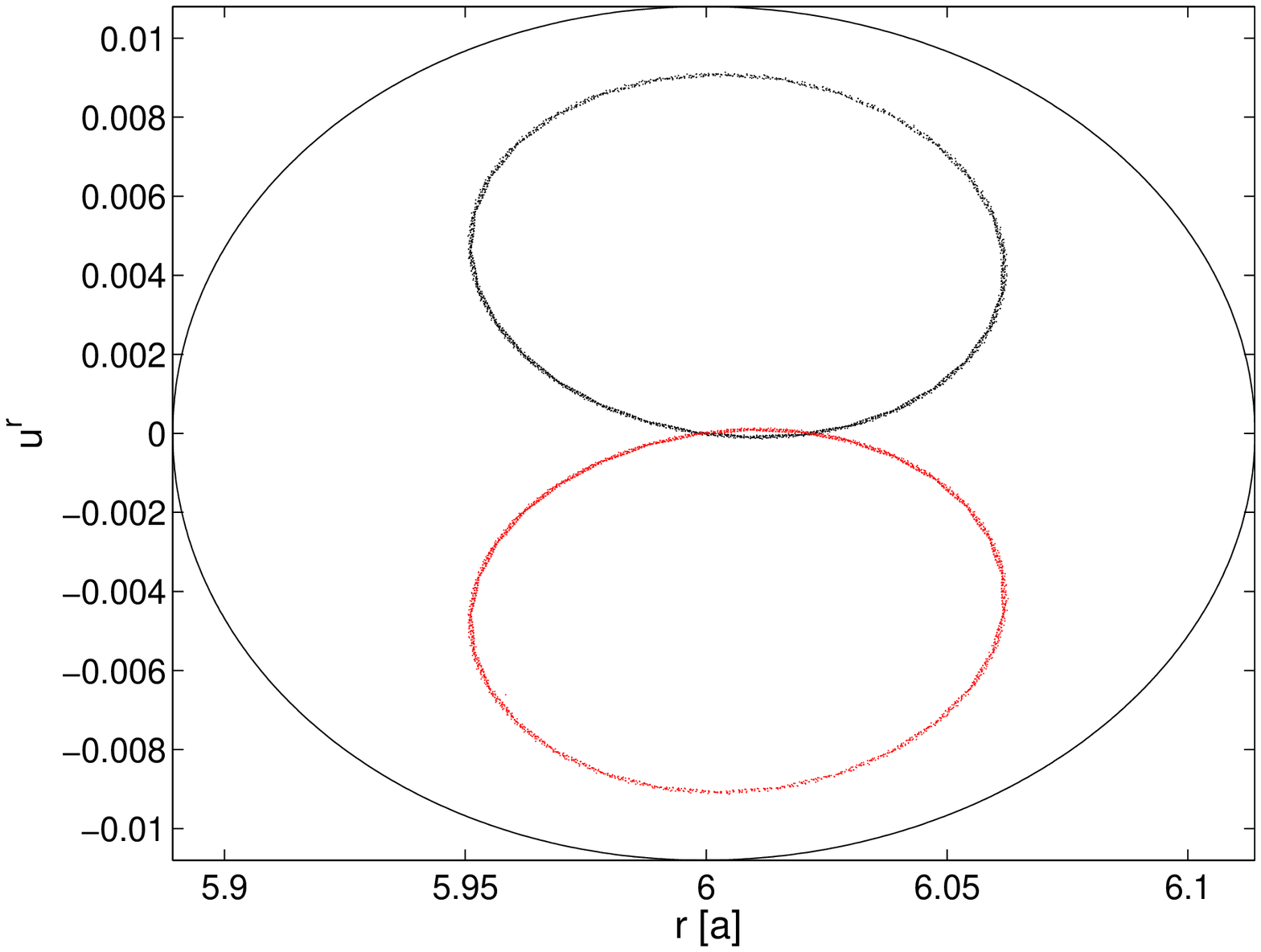}
\includegraphics[scale=0.18,trim=0mm 0mm 0mm 0mm,clip]{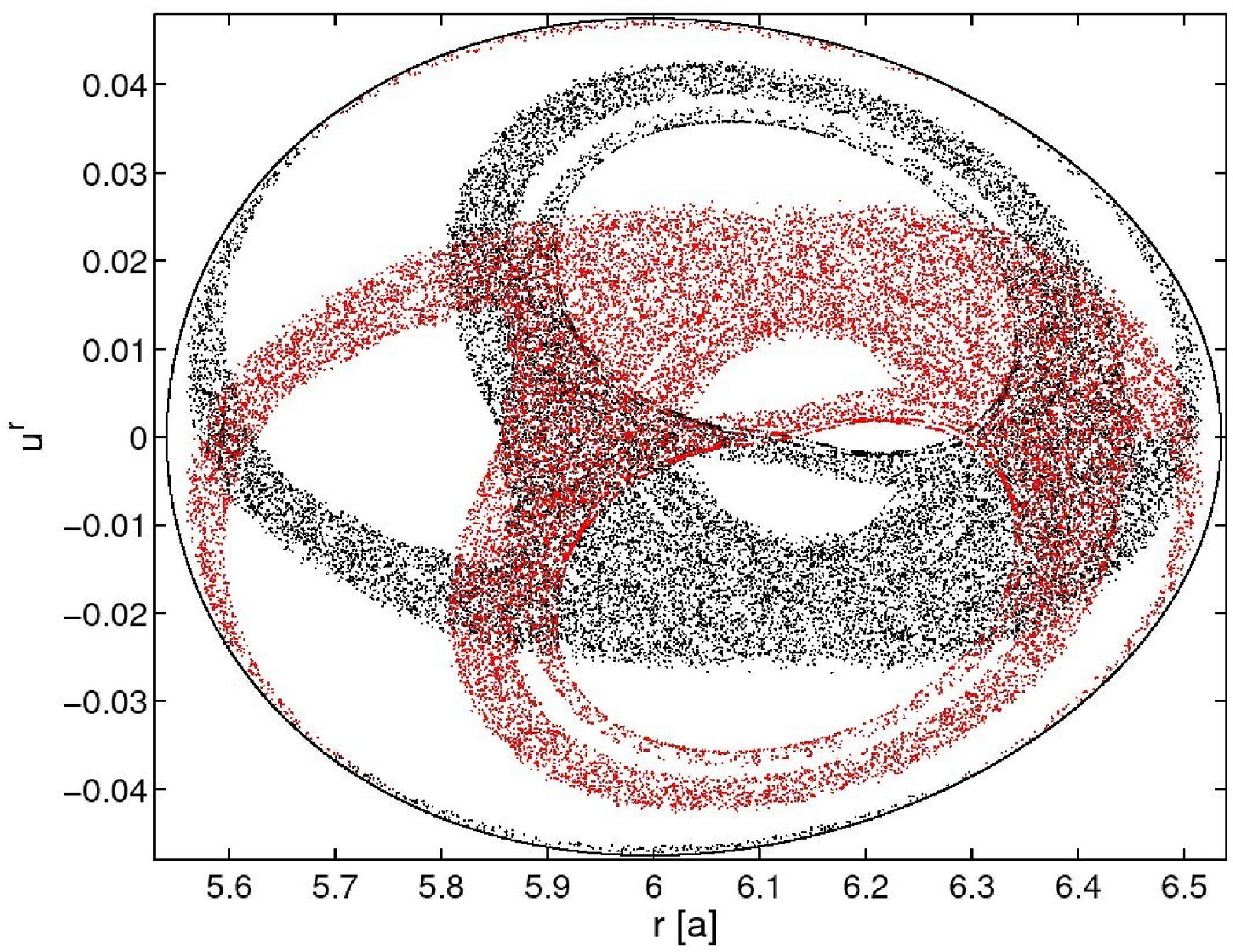}
\includegraphics[scale=0.18,trim=0mm 0mm 0mm 0mm,clip]{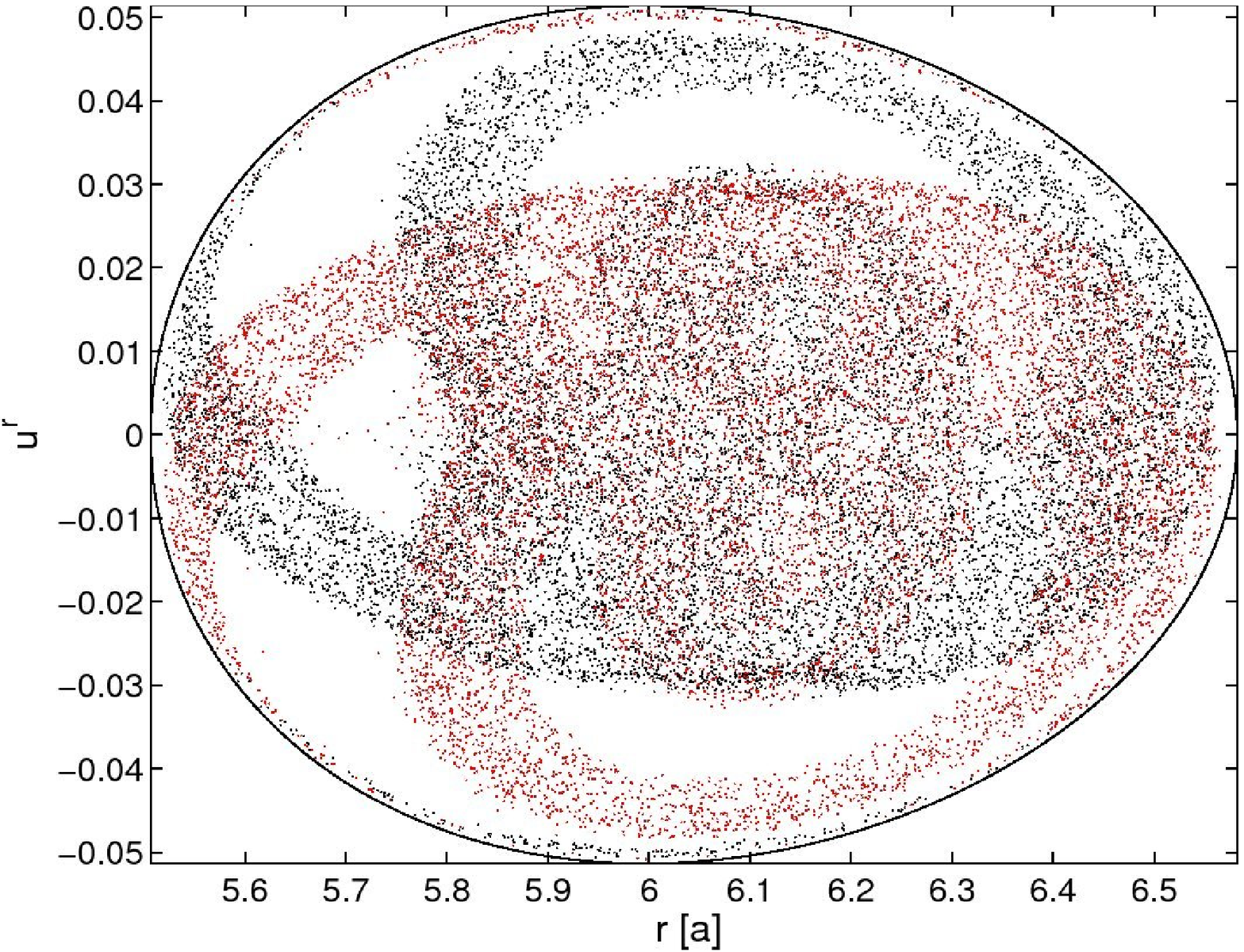}\\
\includegraphics[scale=0.2,trim=0mm 0mm 0mm 0mm,clip]{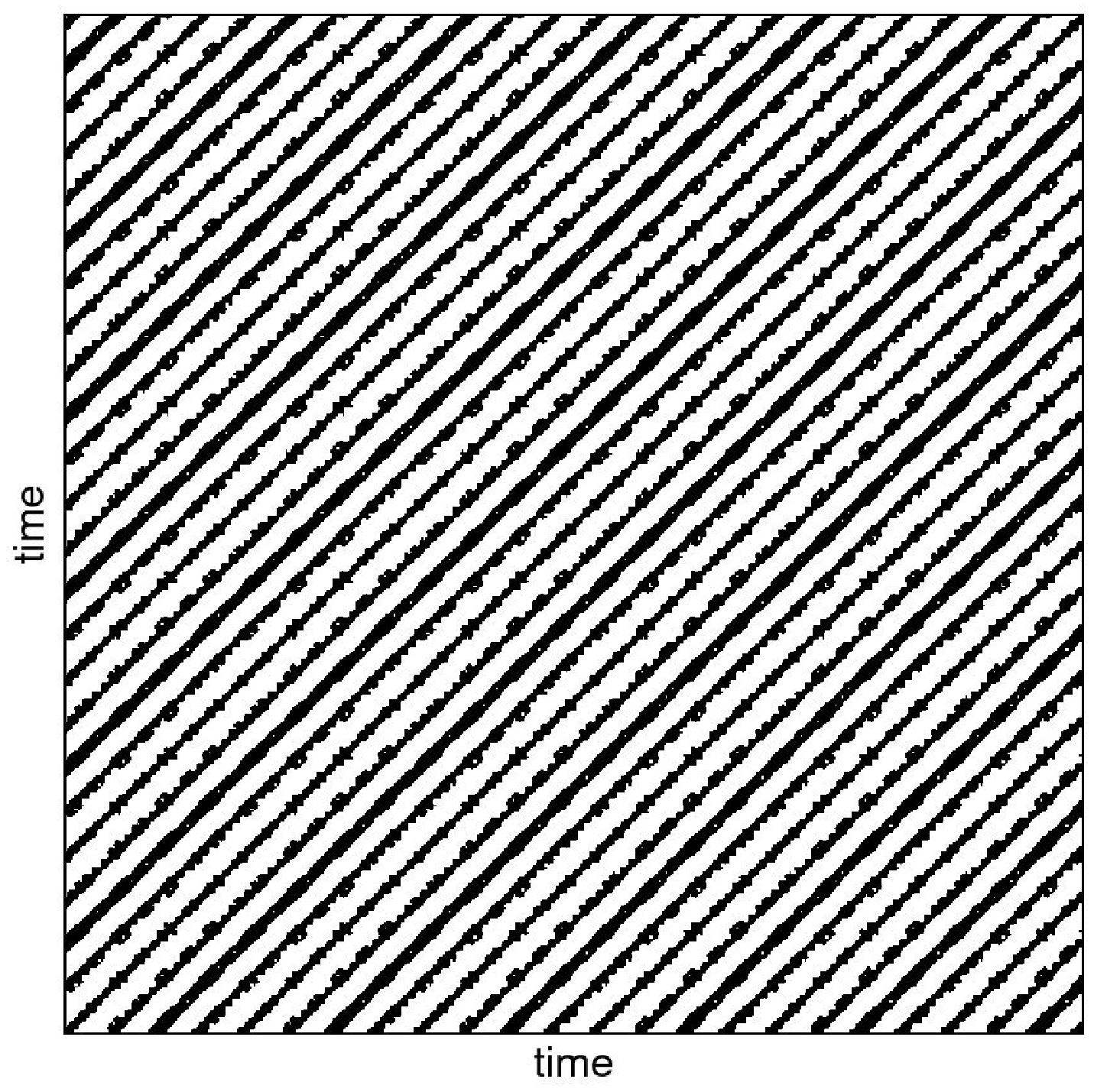}~~~
\includegraphics[scale=0.21,trim=0mm 0mm 0mm 0mm,clip]{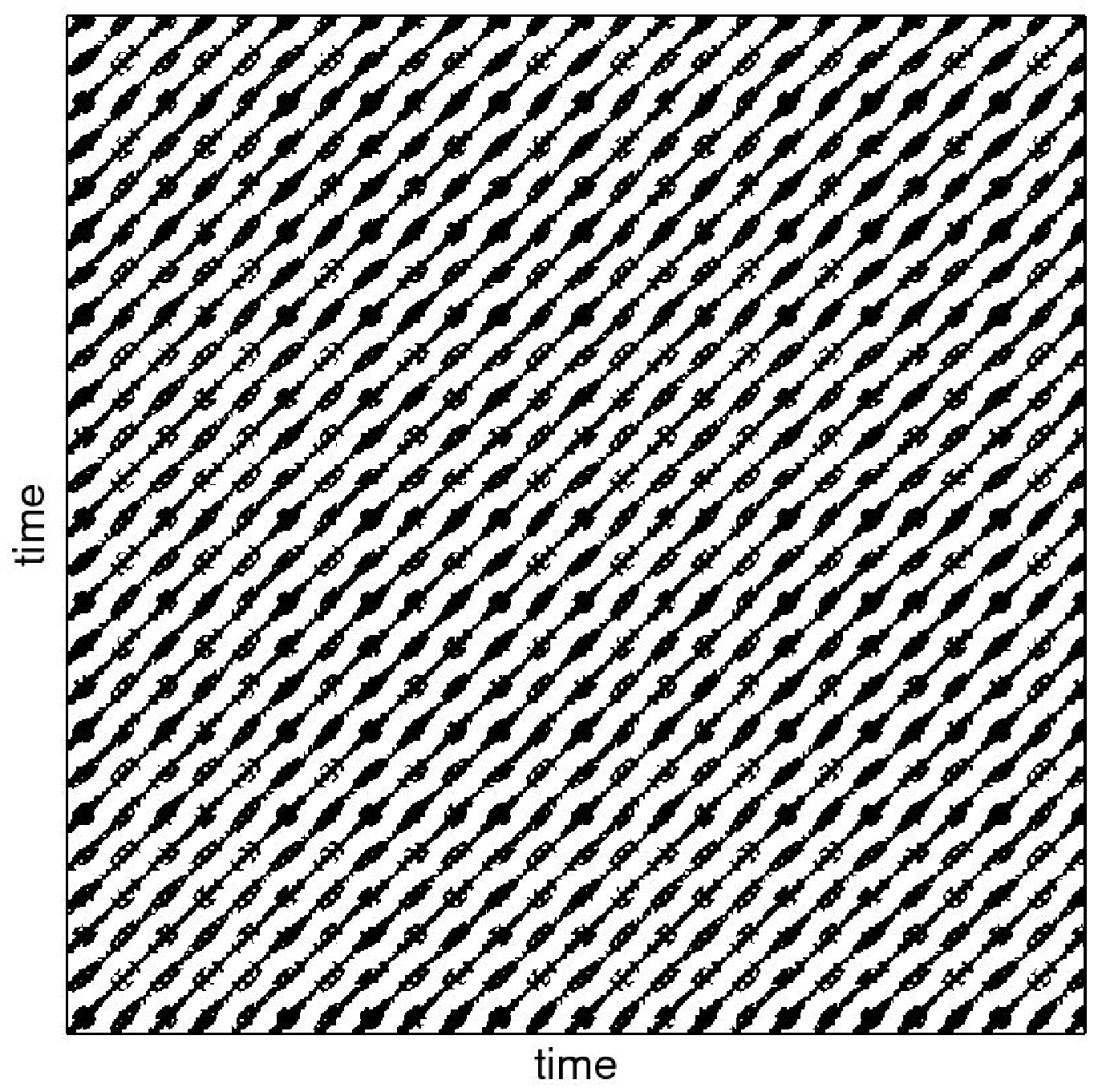}~~~
\includegraphics[scale=0.19,trim=0mm 0mm 0mm 0mm,clip]{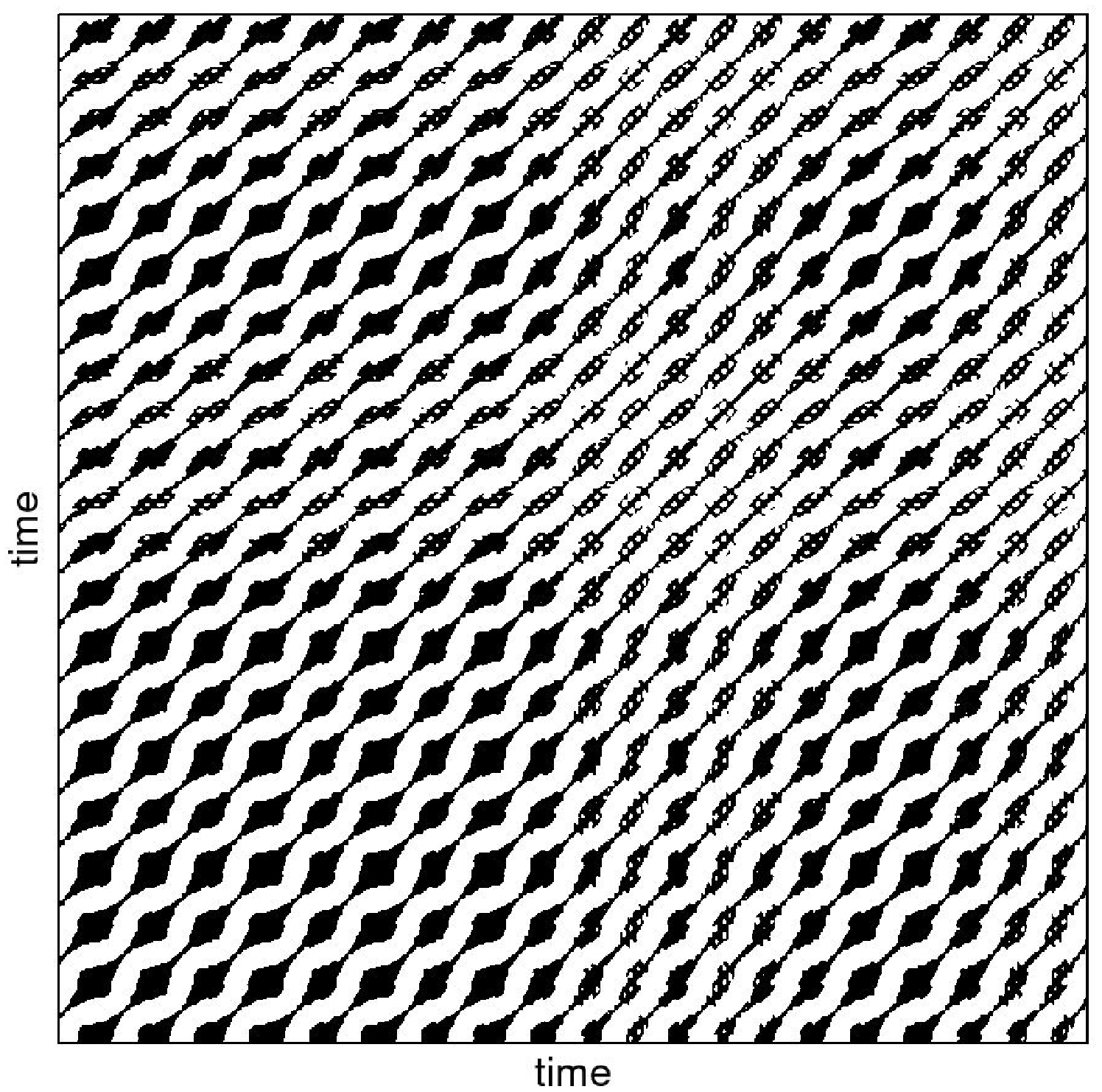}\\
\includegraphics[scale=0.18,trim=0mm 0mm 0mm 0mm,clip]{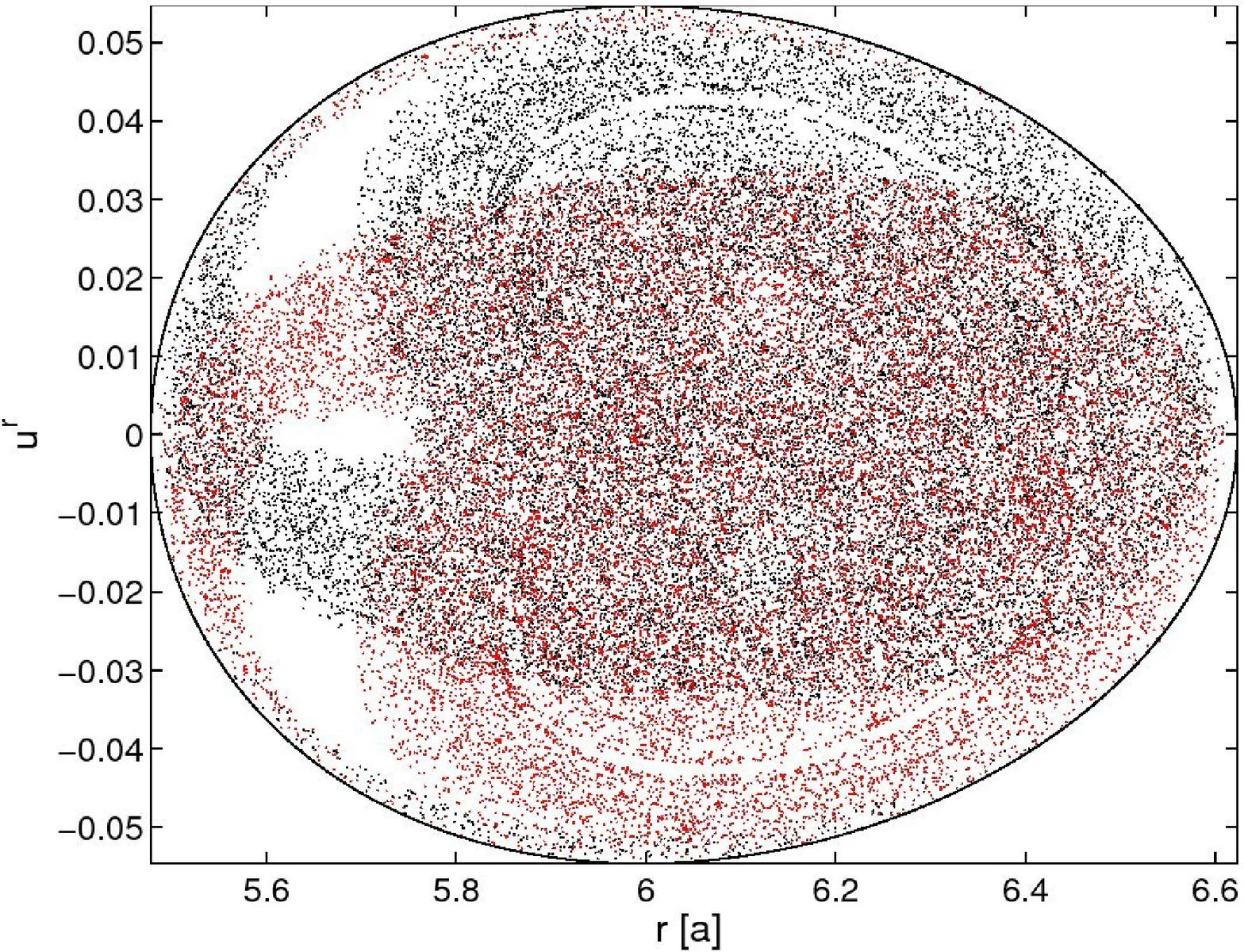}
\includegraphics[scale=0.18,trim=0mm 0mm 0mm 0mm,clip]{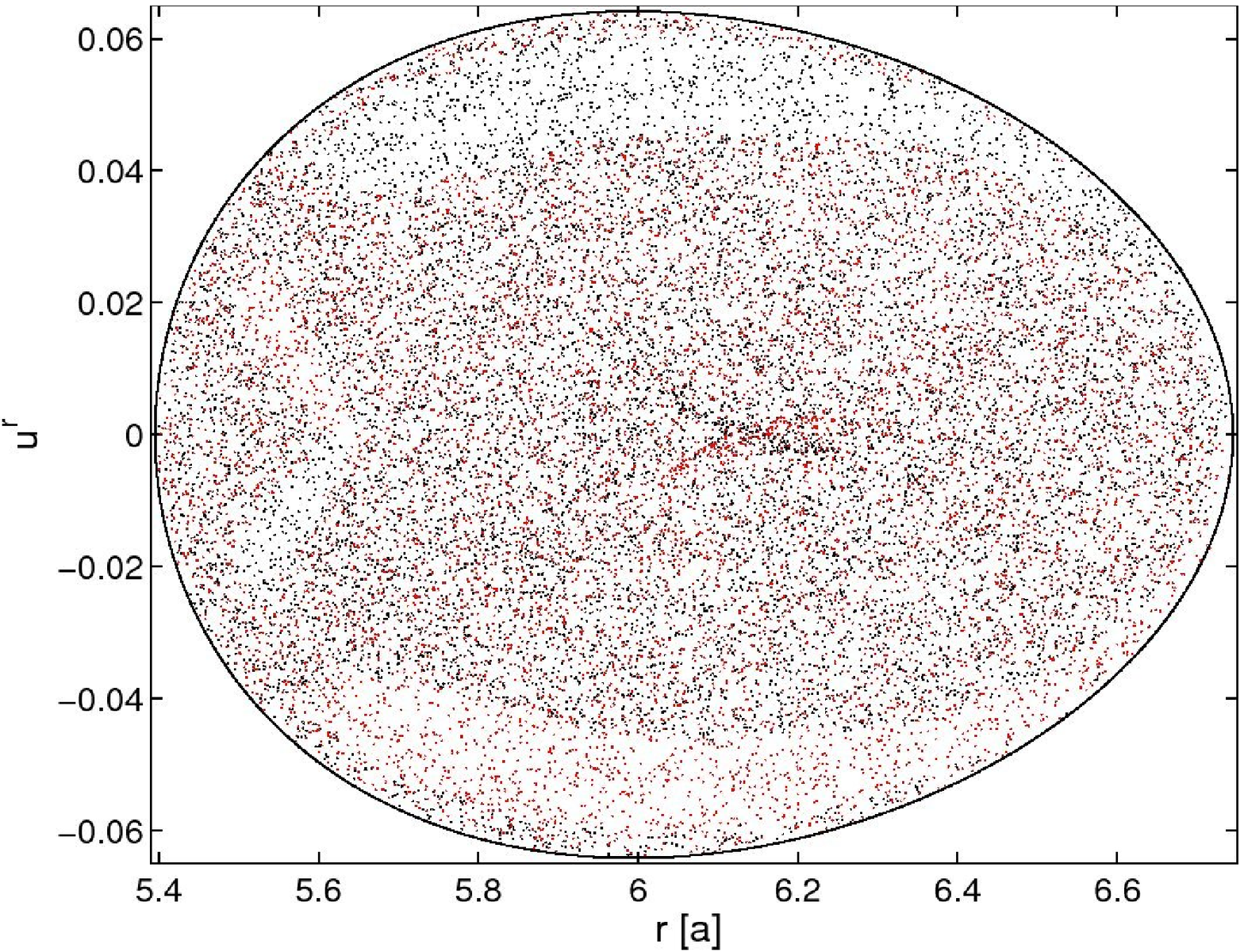}
\includegraphics[scale=0.18,trim=0mm 0mm 0mm 0mm,clip]{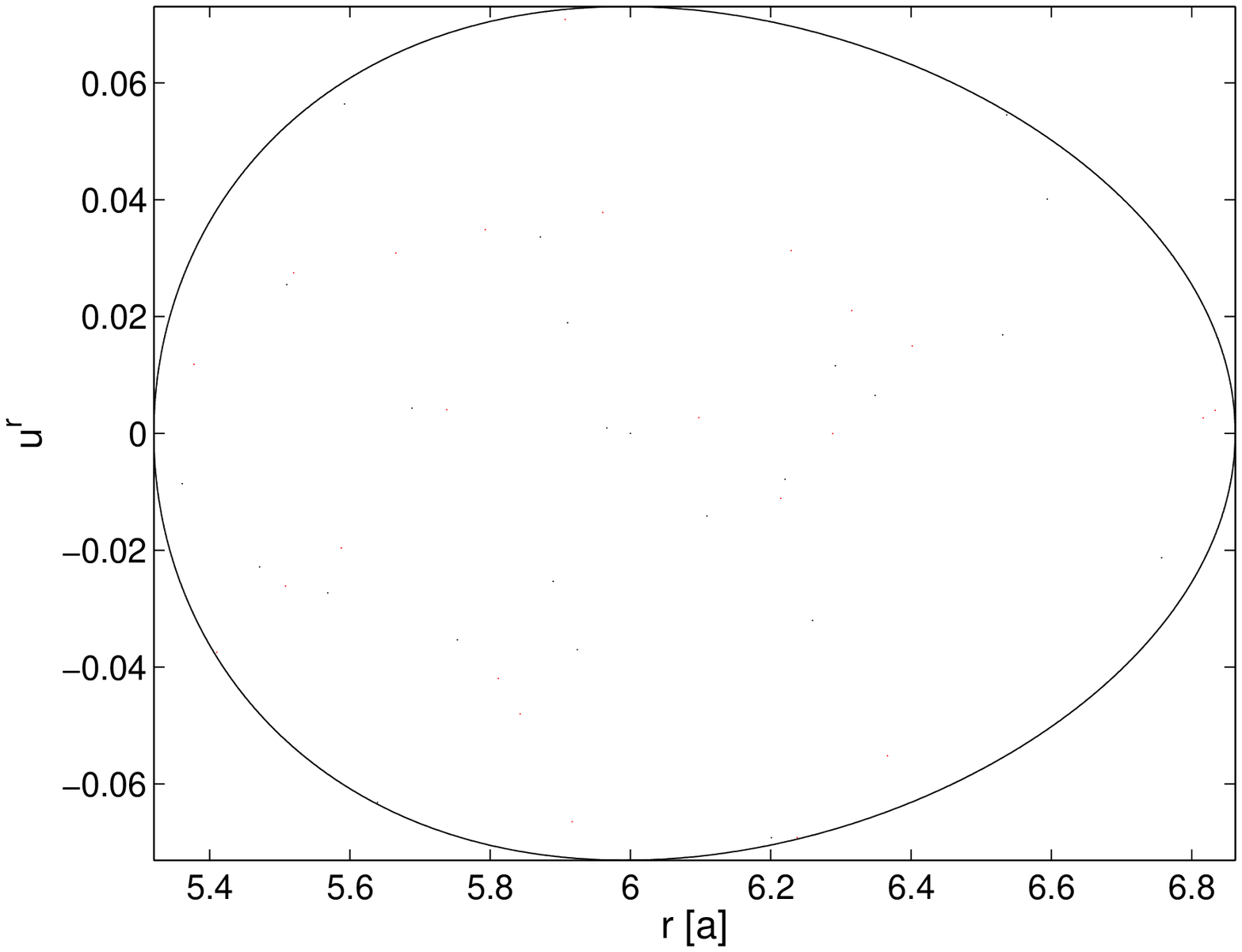}\\
\includegraphics[scale=0.19,trim=0mm 0mm 0mm 0mm,clip]{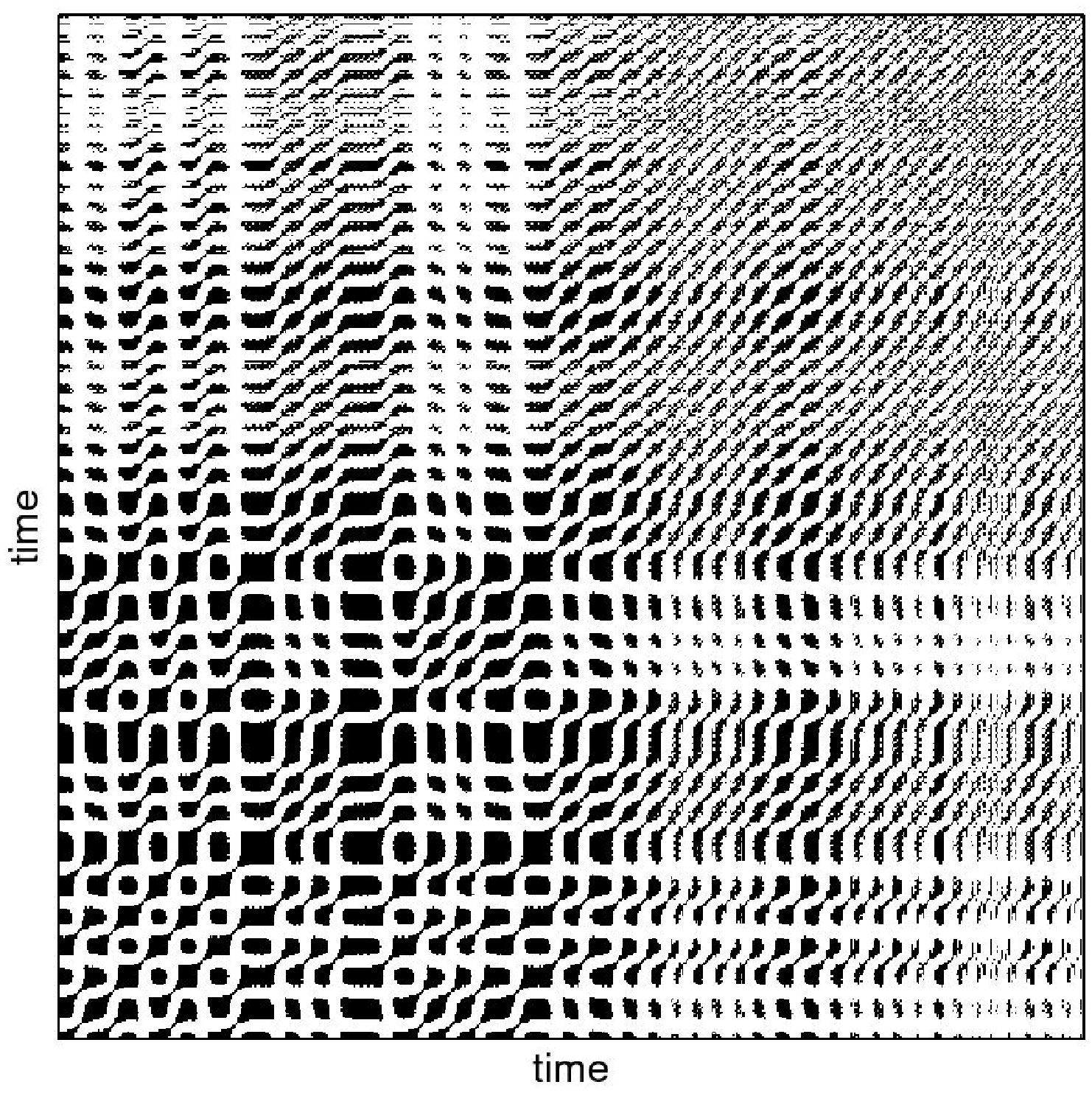}~~~
\includegraphics[scale=0.195,trim=0mm 0mm 0mm 0mm,clip]{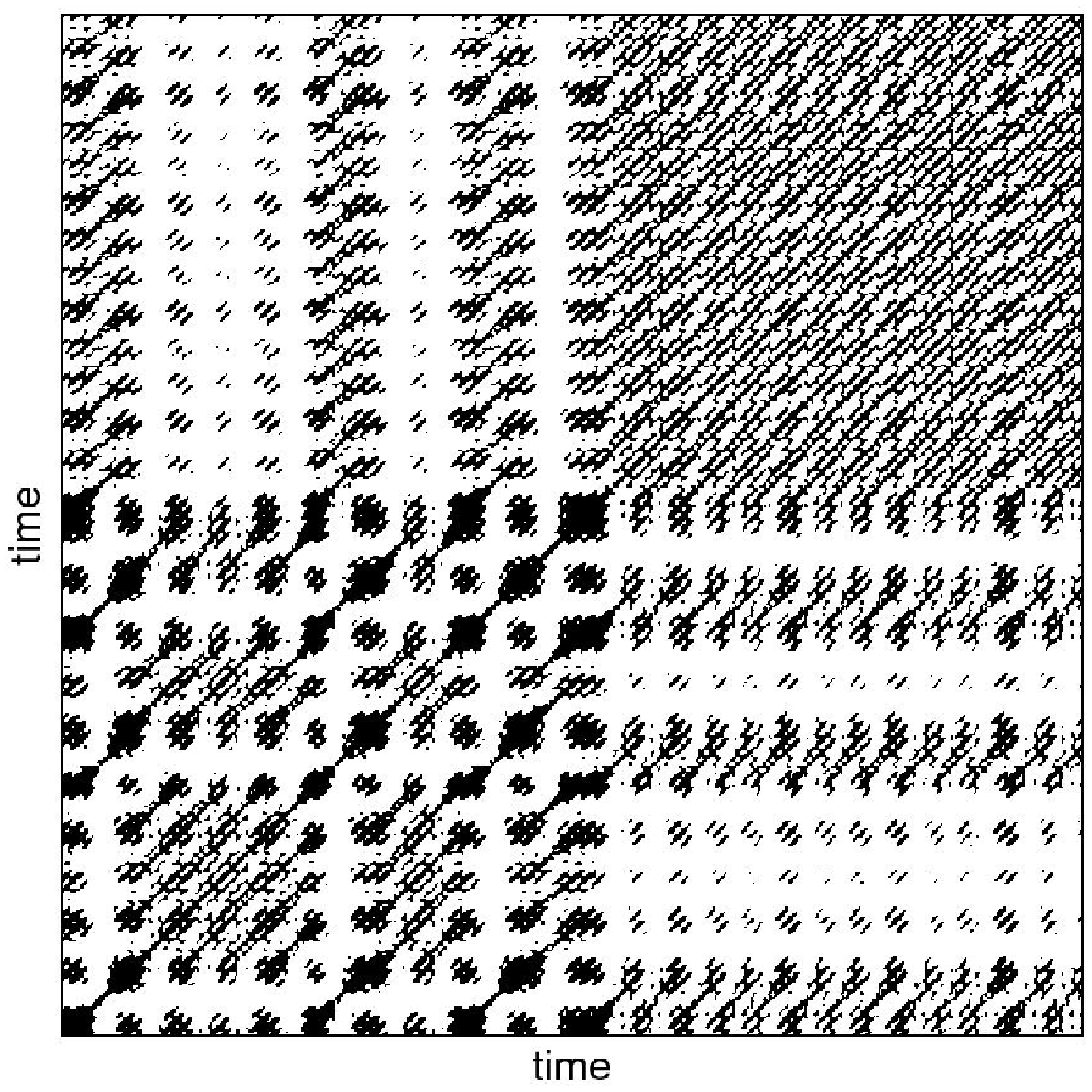}~~~
\includegraphics[scale=0.185,trim=0mm 0mm 0mm 0mm,clip]{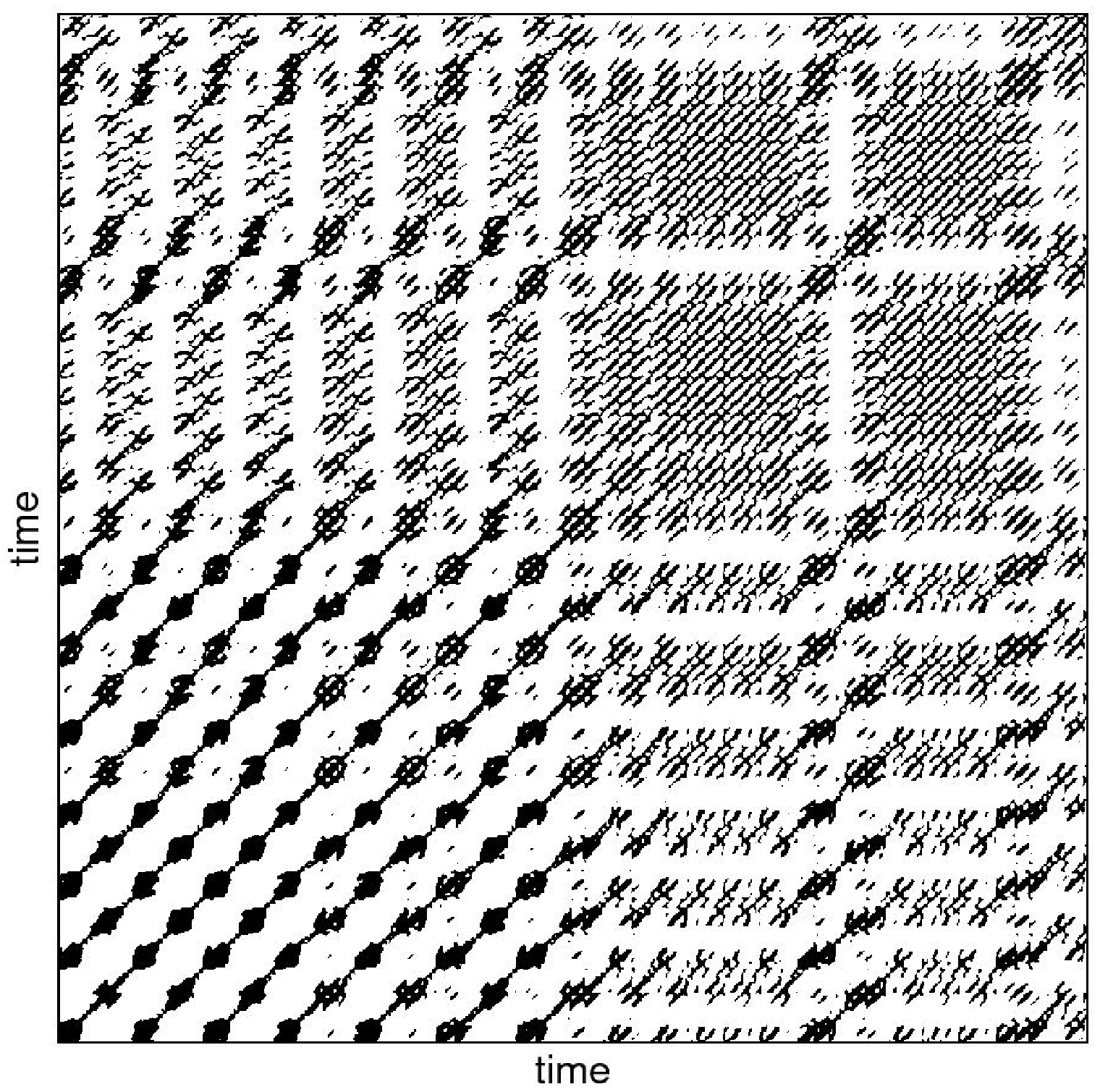}
\end{center}
\caption{Poincar\'{e} surfaces of section ($\theta_{\rm sec}=\pi/3$) and corresponding recurrence plots for charged particle with $q=5.581$ moving within halo lobes in the Bonnor spacetime with $b=1\,a$. Particle with $L=-2.356\,a$ is launched just from the locus of the off-equatorial potential well minimum ($r_{\rm min}=6\,a$, $\theta_{\rm min}=\theta_{\rm sec}=\pi/3$ and $V_{\rm min}=0.81675$) with various values of energy (from the upper left to the bottom right:  $E=0.8168$, $E=0.818$, $E=0.8182$, $E=0.8183$, $E=0.819$ and $E=0.8198$). First three pairs of plots show the situation in the halo lobe, while bottom plots reveal the dynamics after merging of the lobes. A decisive surface of section cannot be constructed for the particle in an opened lobe ($E=0.8198$) as it escapes after several intersections with the surface, while the corresponding RP shows the chaotic nature of the motion unambiguously. Unlike the previous figures, we plot all intersection points: downward crossings with $u^{\theta}\geq0$ (black dot) as well as those resulting from upward crossings with $u^{\theta}<0$ (red dot) on the Poincar\'e surfaces.}
\label{Fig:13}
\end{figure}

If we perturb the system by the magnetic field, however, the chain of Birkhoff islands develops. Although we know that there are some integrable system with resonant islands of single multiplicity \cite{contopoulos02}, here its presence arouses suspicion of nonintegrability, since none were present for $b=0$. Indeed, in the following (see figure \ref{Fig:11}), we observe chaotic motion in this setup ($q=0$, $b\neq0$), which is irrefutable evidence of being nonintegrable. Finally, in the bottom panel of figure \ref{Fig:10}, we introduce the charge of the test particle. We choose such a combination of parameters that leads to the same value of ratio $r_{\rm min}/r_{\rm h}$, as it acquired in the previous uncharged case. This makes the two cases better comparable and the effect of the newly introduced electromagnetic forces more distinguishable. In the last surface of section, we really observe much more complex patterns compared to those of uncharged particles. KAM curves of quasiperiodic orbits are present as well as several Birkhoff chains of islands corresponding to the resonances of intrinsic frequencies of the system. These are interwoven with pronounced chaotic layers. Such a picture is typical for a considerably perturbed system far from integrability. 

However, further examination of the dynamics in opened potential lobes in the non-magnetized system presented in figure \ref{Fig:10b} reveals presence of narrow zones of chaotic orbits which proves the system nonintegrable. These chaotic orbits correspond to those particles which actually leave the potential well after certain amount of time. Coincidentally, the nonintegrability of motion in a general Zipoy-Voorhees spacetime was very recently shown by Lukes-Gerakopoulos in \cite{lukes12} where the issue was treated in detail. We summarize that although the dynamics in the non-magnetized Bonnor spacetime (i.e. Zipoy-Voorhees with $\delta=2$) is typically regular (we actually found no chaotic orbit in closed lobes) the underlying system is not integrable. 

In the following, we compare dynamics in off-equatorial potential wells for uncharged (figure \ref{Fig:11}) and charged particles (figure \ref{Fig:12}). For the sake of better comparability, these are both chosen to have $\theta_{\rm min}=\theta_{\rm sec}=\pi/3$ and equal value of $r_{\rm min}/r_h$. Both series show Poincar\'{e} surfaces of section of particles being launched from the vicinity of off-equatorial potential minima (in which the circular halo orbit resides) differing in energy $E$, which governs the size and shape of the lobe ($E$ sets the level at which the effective potential surface is being intersected). Comparing figures \ref{Fig:11} and \ref{Fig:12} we conclude that charging the test particle makes it more prone to chaotic dynamics and we also observe that the energy of the particle triggers chaotic motion. 
 
To illustrate the continuous transition from ordered to chaotic dynamics, we pick a particular trajectory of charged particle and plot series of Poincar\'e sections along with corresponding recurrence plots in figure \ref{Fig:13}. We launch the particle from the locus of the off-equatorial potential minimum with various values of energy $E$, while other parameters remain fixed. The sequence begins with the energy corresponding to a small halo lobe where we observe ordered motion manifested by narrow curves on the surface of section and, simple diagonal pattern of the recurrence plot (upper left panels of figure \ref{Fig:13}).  Increasing the energy, however, gradually shifts the dynamics towards deterministic chaos -- trajectory becomes more and more ergodic as it spans larger fraction of given energy hypersurface in the phase space.

\section{Conclusions}
In this contribution  we presented a brief numerical study of test particle dynamics occurring in the Bonnor spacetime. First we analysed motion in equatorial potential wells in three different cases, namely motion in non-magnetized spacetime with $b=0$, motion of uncharged particles on the magnetized background ($q=0$, $b\neq 0$) and dynamics in the general case $q\neq 0$, $b\neq 0$. Our results show that without magnetic field the system hosts mostly regular orbits and the dynamics of test particles resembles closely fully integrable systems. However, further numerical inspection revealed that chaotic orbits are also present in this setup proving the system nonintegrable. Then we observed that the magnetic parameter $b$ introduces profound perturbation of the dynamics. Moreover, a charge of particle acts as an extra perturbation, which shifts magnetized system even farther from the integrability.

Within the off-equatorial potential wells we also studied the role of particle energy $E$ on the degree of chaos found in the system, concluding that it acts as a trigger for the chaotic motion. As the energy is gradually increased, the system undergoes a continuous transition from the regular behaviour to the chaotic dynamics, being almost fully ergodic on the given hypersurface. We illustrated such transition by means of Poincar\'e surfaces of section and recurrence plots. 

\section*{Acknowledgements}
Authors appreciate support from the following projects: GA \v{C}R ref.\ 202/09/0772 (OK), GA \v{C}R ref.\ P209/10/P190 and Synergy CZ.1.07/2.3.00/20.0071 (JK) and Czech-US collaboration project ME09036 (VK).

\section*{References}
\bibliography{ae100prg}

\end{document}